\documentclass
[aps,prc,twocolumn,showpacs,showkeys,amsmath,floatfix,superscriptaddress]{revtex4}
\usepackage{color,graphicx,xcolor}
\usepackage[utf8]{inputenc}
\usepackage{mathptmx}                
\usepackage{dcolumn}                 
\usepackage{bm}                      

\begin{document}
\title{Thermodynamics of baryonic matter with strangeness \\
within non-relativistic energy density functional models}
\author{Ad. R. Raduta}
\affiliation{IFIN-HH, Bucharest-Magurele, POB-MG6, Romania}

\author{F. Gulminelli}
\affiliation{ENSICAEN, UMR6534, LPC ,F-14050 Caen c\'edex, France}

\author{M. Oertel}
\affiliation{LUTH, CNRS, Observatoire de Paris, Universit\'e Paris Diderot, 
5 place Jules Janssen, 92195 Meudon, France}

\begin{abstract}
  We study the thermodynamical properties of compressed
  baryonic matter with strangeness within non-relativistic energy density
  functional models with a particular emphasis on possible phase transitions
  found earlier for a simple $n,p,e,\Lambda$-mixture. The aim of the paper is
  twofold: I) examining the phase structure of the complete system, including
  the full baryonic octet and II) testing the sensitivity of the results to
  the model parameters.  We find that, associated to the onset of the
  different hyperonic families, up to three separate strangeness-driven phase
  transitions may occur. Consequently, a large fraction of the baryonic
  density domain is covered by phase coexistence with potential relevance for
  (proto)-neutron star evolution.  It is shown that the presence of a phase
  transition is compatible both with the observational constraint on the
  maximal neutron star mass, and with the present experimental information on
  hypernuclei.  In particular we show that two solar mass neutron stars are
  compatible with important hyperon content.  Still, the parameter space is
  too large to give a definitive conclusion of the possible occurrence of a
  strangeness driven phase transition, and further constraints from
  multiple-hyperon nuclei and/or hyperon diffusion data are needed.
 
\end{abstract}

\date{\today}


\maketitle

\section{Introduction}
\label{section:intro}

In the effort of building more realistic equations of state (EoS) on which the
understanding of astrophysical issues as the structure and evolution of
neutron stars (NS) or core-collapsing supernovae (CCSN) relies, special
attention is presently paid to the behavior of baryonic matter at densities
above nuclear matter saturation density.  The subject is challenging as
experimental data are too scarce to satisfactorily constrain the respective
interactions, in particular if non-nucleonic degrees
of freedom are involved.

Though, simple energetic arguments show that no reliable description can be
conceived without considering strangeness~\cite{glendenning}.  As such it is
hoped that astrophysical observations can eventually supplement the missing
knowledge so far attained in terrestrial laboratories.  An example in this
sense is the present debate about the measurement of very massive neutron
stars, and the associated core composition.  The early conclusions ruling out
hyperons from the NS core seem to be refuted by recent relativistic and
non-relativistic mean-field models showing that a sufficiently repulsive
hyperon-nucleon ($Y$-$N$) and hyperon-hyperon ($Y$-$Y$) interaction at high
densities is able to reconcile the two solar mass measurements corresponding
to PSR J1614-2230~\cite{demorest} and PSR J0348+0432~\cite{antoniadis} with
the onset of
strangeness~\cite{Bednarek11,micaela_prc85,sedrakian_aa2012,weissenborn11a,weissenborn11b}
without necessarily a very early deconfinement transition circumventing the
hyperon puzzle~\cite{Zdunik_aa13}.  The presence of hyperons in dense stellar
matter is expected to have important astrophysical consequences.
We can recall for instance the modification of the neutron star cooling rate due to
hyperonic Urca processes~\cite{haensel_94, schaab_96} leading to very fast
cooling for stars with a mass high enough to allow for the onset of hyperons,
a result, however, very sensitive to hyperonic pairing~\cite{cooling}, and
thus subject to large uncertainties.
By allowing for weak non-leptonic reactions ($N+N \leftrightarrow N+Y$, 
$N+Y \leftrightarrow Y+Y$), direct and modified hyperonic Urca and strong
interactions ($Y+Y \leftrightarrow N+Y$, $Y+Y \leftrightarrow Y+Y$)
hyperons are also shown to impact on bulk viscosity and, thus, damp
r-mode instabilities \cite{r-mode}.
 
Most of the predictions are done within mean-field models. However, because of
generic attractive and repulsive couplings between the different baryonic
species, phase transitions could, in principle, be faced. An example is the
liquid-gas phase transition occurring in nuclear matter. If there is a phase
transition the mean-field solutions should be replaced by the Gibbs
construction in the phase coexisting domains, thus modifying the equation of
state.  The occurrence of a phase transition in strange compressed baryonic
matter has already been discussed in Ref.~\cite{schaffner-bielich_2002}, where a
new family of neutron stars characterized by much smaller radii than usually
considered was predicted.  However, very attractive hyperon-hyperon couplings
were considered in that study, which presently appear ruled out by the
experimental information on the ground state energy of double-lambda
hypernuclei.

A detailed study of the phase diagram of dense baryonic matter was recently
undertaken in Ref.~\cite{nL,npLe} within a non-relativistic mean-field model
based on phenomenological functionals. The models in Refs.~\cite{nL,npLe}
considered a simplified setup, taking only $(n,\Lambda)$~\cite{nL} and
$(n,p,\Lambda)$~\cite{npLe} baryon mixtures into account. It was shown that
under these assumptions first- and second- order phase transitions exist, and
are expected to be explored under the strangeness equilibrium condition
characteristic of stellar matter.  Two astrophysically relevant consequences
have been worked out. In Ref.~\cite{npLe} it has been demonstrated that in the
vicinity of critical points the neutrino mean-free path is dramatically
reduced, such that the neutrino transport can be considerably affected.
Ref.~\cite{Peres_2013} shows within a spherical simulation that, if during
the proto-neutron star contraction after bounce the phase coexistence region
is reached, a mini-collapse is induced, leading to pronounced oscillations of
the proto-neutron star.  

Our previous work suffers nevertheless of two major limitations.  First, we
considered the strangeness degree of freedom as fully exhausted by the
$\Lambda$ hyperon. If the hyperonic couplings are such that $\Sigma^-$ or
$\Xi^-$ are more abundant than $\Lambda$, as predicted by a series of models,
the extension and localization of phase coexistence domains will be
different. Moreover, the possible dominance of a charged hyperon would impact
on the direction of the order parameter and change the stability of the phase
diagram with respect to the electron gas. 
In the extreme case, it could
even make it disappear for neutron star matter.

Secondly, the previous study~\cite{nL,npLe} employs only one specific model,
a phenomenological energy density functional, with one parameter set,
assuming in particular a strongly attractive hyperon-hyperon interaction. 
The most recent analysis
\cite{Aoki_2009,ahn_2013} of double-$\Lambda$ hypernuclei now tend to suggest
a very small attraction in the $\Lambda$-$\Lambda$ channel, though experimental
data are still very scarce and extrapolations to infinite matter
uncertain. Moreover accurate fits exist of microscopic Brueckner
calculations~\cite{BSL} providing functionals which are consistent with the
available experimental data on nucleon-$\Lambda$ phase shifts. 
This can give some guidance on the functional 
in the
nucleon-$\Lambda$ channel, although it is known that such models fail to
reproduce the existence of very massive neutron stars.

The aim of the present paper is thus twofold.  Firstly, to investigate the
phase diagram of the whole baryonic octet.  Secondly, to test the
sensitivity of the strangeness-driven phase transition on the hyperonic
coupling constants.  Both $NY$ and $YY$ channels will be considered.  Since
most experimental constraints concern the $\Lambda$-hyperon, to avoid
proliferation of uncontrolled parameters, we will consider for this study only
the simplest mixture $(n, p, \Lambda)+e$ which accounts for the three relevant
densities: baryonic, charge/leptonic and strangeness.

The paper is organized as follows.  Section~\ref{section:themodel} briefly
presents the model and the phenomenology of the phase transition. The phase
diagram of purely baryonic, as well as charge-neutral baryonic matter with
leptons are spotted in
Sections~\ref{section:uncharge} and~\ref{section:neutral}.  The model
dependence of the results is analyzed in Section~\ref{section:modeldep} by
considering alternative density functionals for the $N$-$Y$ and
$Y$-$Y$-interactions and various values for the coupling constants.
Conclusions are drawn in Section~\ref{section:concl}.

\section{The model}
\label{section:themodel}

In non-relativistic mean-field models, the total energy density 
of homogeneous baryonic matter is
given by the sum of mass and kinetic energy densities of different particle
species and the potential energy density:
\begin{eqnarray}
e_B \left(\{n_i \} \right)&=&
\sum_{i=n,p,\Lambda,\vec \Sigma,\vec \Xi} \left(n_i m_i c^2+\frac{\hbar^2}{2 m_i} \tau_i \right)
+e_{pot}(\{n_i\}) \nonumber \\
&=&e_B \left( n_B, n_S, n_Q \right),
\end{eqnarray}
where $n_B=\sum_i n_i B_i$; $n_S=\sum_i n_i S_i$; $n_Q=\sum_i n_i Q_i$
represent the baryon, strangeness and charge number densities, respectively,
corresponding to the three good quantum numbers assuming equilibrium with respect
to strong interaction.  The particle and kinetic energy densities can be
expressed in terms of Fermi-Dirac integrals,
\begin{equation}
n_i
=\frac{1 }{2 \pi^2 \hbar^3} \left(\frac{2\,m_i}{\beta} \right)^{\frac 32} 
F_{\frac 12}(\beta \tilde \mu_i)  ; \;
\tau_i=\frac{1 }{2 \pi^2 \hbar^5} \left( \frac{2\,m_i}{\beta } \right )^{\frac 52}
F_{\frac 32}(\beta \tilde \mu_i),
\end{equation}
with $F_{\nu}(\eta)=\int_0^{\infty} dx \frac{x^{\nu}}{1+\exp \left(
    x-\eta\right)}$.  $\beta=T^{-1}$, $m_i$ and $\tilde \mu_i$ denote,
respectively, the inverse temperature, the effective $i$-particle mass and the
effective chemical potential of the $i$-species self-defined by the
single-particle density.
The effective chemical potentials are related to the chemical
potentials 
\begin{equation}
\mu_i = \frac{\partial e_B}{\partial n_i} 
\end{equation}
via
\begin{equation}
\tilde \mu_i = \mu_i - U_i - m_i c^2~,
\end{equation}
where $U_i = \left.\frac{\partial e_{\mathit{pot}}}{\partial
    n_i}\right|_{n_j,j\neq i}$ are the
self-consistent mean field single-particle potentials. 

The potential energy density should in principle account for all possible
couplings between nucleonic and hyperonic species, $N$-$N$, $N$-$Y$ and
$Y$-$Y$. The nuclear structure data constrain satisfactorily the
$N$-$N$-interaction up to densities close to the normal nuclear saturation
density and moderate isospin asymmetries, such that well constrained and
reliable expressions for this functional, including isospin dependent
effective masses and currents are available. The situation is much less clear
for higher densities, strong isospin asymmetries as well as for channels
containing hyperons. The most general expression of these potential energies 
can be expanded in a polynomial form
\begin{equation}
e_{pot}(n_C,n_{C'})=\sum_{k,m} a_{CC'}^{(k,m)} n_C^k n_{C'}^{m}
\label{eq:epot_pol}
\end{equation}

As a guideline to characterize the couplings, the
single particle potentials of baryon $C$ in pure $C'$-matter are employed:
$U_C^{(C')}(n_{C'})= \partial e_{pot}(n_C,n_{C'})/\partial n_C|_{n_C=0}$.  
The coupling constants $a_{CC'}^{(k,m)}$ can then be adjusted to reproduce
standard values of these potentials at some reference density, obtained within
a (model dependent) analysis of the available experimental data.  

A Skyrme-like expression has been frequently employed for the energy density, 
where the contribution of channel $CC'$ to the potential energy
density is given by  
\begin{eqnarray}
e_{CC'}( n_C, n_{C'} )=a_{CC'} n_C n_{C'}+ c_{CC'} n_C n_{C'} 
\left( n_C^{\gamma_{CC'}} +n_{C'}^{\gamma_{CC'}} \right);
\nonumber \\
a_{CC'}<0; ~ c_{CC'}>0 ~ \gamma_{CC'}>0.
\label{eq:epot_CC'}
\end{eqnarray}
This form, which depends on only three parameters for each channel, is the
simplest expression which corresponds to a controlled compressibility and
fulfills the condition that $U_C^{(C')}(n_{C'})$ vanishes at vanishing
$C'$-density $\lim_{n_{C'} \to 0} U_C^{(C')} (n_{C'} ) \to 0$, and 
becomes highly repulsive
at $C'$-
high density $\lim_{n_{C'} \to \infty} U_C^{(C')} (n_{C'} ) \to \infty$.
Let us notice that this simple (and probably simplistic!) form
together with the fact that we fix it at one given density
implies a coupling  between short-
and long-range behaviors of $U_C^{(C')}$ and $U_C^{(C)}$ potentials,
$U_C^{(C')}(n_{C'})=a_{C C'} n_{C'}+ b_{C C'} n_{C'}^{\gamma_{C C'}+1}$,
$U_C^{(C)}(n_C)=2 a_{CC} n_C+2 (\gamma_{CC}+1) c_{CC} n_C^{\gamma_{CC}+1}$.

Concerning the channels including strangeness, the available
experimental information is particularly scarce. Hypernuclei
experiments only provide information on $\Lambda-$, $\Sigma-$ and
$\Xi-$ potential well depths in symmetric nuclear matter at saturation
densities and, to a less accurate extent, on the $\Lambda$-$\Lambda$
interaction potential. Based on a wealth of
$\Lambda$-hypernuclear data produced in $(\pi^+,K^+)$ reactions, the
presently accepted value of $U_{\Lambda}^{(N)}(n_0)$ is considered to
be $ \approx -30$ MeV~\cite{UNL}.  $U_{\Xi}^{(N)}(n_0)$ is known to be
attractive, too, $\approx -14$ MeV, based on missing mass measurements
in the $(K^-,K^+)$ reaction on carbon~\cite{UNXi}.  The situation of
$U_{\Sigma}^{(N)}(n_0)$ is ambiguous.  On the one hand $(\pi^-,K^+)$
reactions on medium-to-heavy nuclei point to a repulsive potential of
the order of 100 MeV or less~\cite{UNSigma_repulsive}.  On the other
hand, the observation of a $^4_{\Sigma}$He bound state in a
$^4$He($K^-,\pi^-$) reaction~\cite{UNSigma_attractive} pleads in favor
of an attractive potential.  Very few multi-hyperon exotic nuclei data
exist so far and all of them correspond to double-$\Lambda$ light
nuclei.  The $\Lambda$-$\Lambda$ bond energy
can be estimated from the binding energy difference between
double-$\Lambda$ and single-$\Lambda$ hypernuclei,
\begin{equation}
\Delta B_{\Lambda\Lambda}=B_{\Lambda \Lambda}(^A_{\Lambda \Lambda}Z)-2 B_{\Lambda}(^{A-1}_{\Lambda}Z),
\end{equation}
where
\begin{equation}
B_{\Lambda\Lambda}(^A_{\Lambda \Lambda}Z)=B(^{A}_{\Lambda \Lambda} Z) - 
B(^{A-2} Z).
\end{equation}
Measured bond energies are affected by huge error bars.  Double-Lambda
$^{10}_{\Lambda \Lambda}$Be and $^{13}_{\Lambda \Lambda}$B data suggest
$\Delta B_{\Lambda\Lambda}\approx 5$ MeV \cite{exp_2lambda} while
$^{6}_{\Lambda \Lambda}$He data point toward a lower value $\Delta
B_{\Lambda\Lambda}= 0.67 \pm 0.17 $ MeV \cite{Aoki_2009,ahn_2013}.  The bond
energy can be interpreted as a rough estimation of the
$U_{\Lambda}^{(\Lambda)}$ potential at the average $\Lambda$ density $\langle
n_\Lambda\rangle$ inside the hypernucleus~\cite{vidana_2001}.  The
extrapolation of few body systems binding energies into mean-field quantities
is, nevertheless, problematic.  In this sense, the more attractive values
extracted from larger nuclei could appear more appealing for the present
application.  Though, it is presently considered that the most accurate
experimental data correspond to the Nagara event with a bond energy of$
0.67\pm 0.17$ MeV.  In this work we shall therefore consider the latter value,
$\Delta B_{\Lambda\Lambda}= 0.67$ MeV.  In what regards the average
$\Lambda$-density in light nuclei we shall use the value proposed in
Ref. \cite{vidana_2001}, $\langle n_\Lambda\rangle\approx n_0/5$.
  
It is clear that fixing the standard value of the
potential at one specific density only constrains one parameter of the
generic $C$-$C'$-interaction. This shows that any phenomenological
mean-field parameterization is subject to large uncertainties.

In our previous studies~\cite{nL,npLe} we have employed the Skyrme-based f
unctional by  Balberg and Gal~\cite{BG97},
\begin{eqnarray}
e_{\mathit{pot}}^{(BG)}\left( \{ n_i\} \right)&=&\sum_{i,j} e_{ij}^{(BG)} (n_i,n_j); \nonumber \\
e_{ij}^{(BG)} (n_i,n_j)&=&\left( 1-\frac{\delta_{ij}}{2} \right)
( a_{ij} n_i n_j +b_{ij}  n_{i3} n_{j3} \nonumber \\ 
&+&c_{ij} \frac{n_i^{\gamma_{ij}+1} n_j + n_j^{\gamma_{ij}+1} n_i}{n_i+n_j}
),
\label{eq:epot_BG}
\end{eqnarray} 
where $n_i,n_j$ are the isoscalar densities for nucleons, and $\Lambda$-,
$\Sigma$- and $\Xi$-hyperons.  $n_{i3}$ stands for the respective iso-vector
densities and the values of $\gamma_{ij}=: \gamma $ are chosen identical for
any $(i,j)$ for simplicity.  As one may notice, the same functional form is
employed in all channels and the potential energy proposed by
Eq.~(\ref{eq:epot_BG}) deviates from the simple polynomial form of
Eq.~(\ref{eq:epot_CC'}) truncated at low order
because of the $1/(n_i+n_j)$-dependence of the
short-range term.  
The expressions of the single-particle potentials of baryon $C$
in pure $C$- and, respectively, $C'$-matter are
\begin{eqnarray}
U_C^{(C)}(n_{C})= a_{CC} n_{C}+\frac{\gamma+1}2 c_{CC} n_{C}^{\gamma} , 
\nonumber \\
U_C^{(C')}(n_{C'})=a_{CC'} n_{C'}+c_{CC'} n_{C'}^{\gamma}~.
\label{eq:coupling}
\end{eqnarray}
They show that fixing the potentials, introduces a correlation between the
  parameters governing repulsion and those governing attraction not present in
  the generic form of Eq. (\ref{eq:epot_pol}).  As we will see, such a
correlation is also present in BHF microscopically based energy functionals
proposed in Ref. \cite{BSL}.
A deeper anlaysis with a larger number of models would, however, be necessary to
check whether this particularity possibly affects the generality of our
results. A study in this sense with relativistic models is in progress
\cite{micaela_rmf}.

Ref.~\cite{BG97} proposes three sets of parameters
corresponding to different stiffnesses $\gamma=2,5/3,4/3$.  For the sake of
simplicity, a unique value is assumed for $a_{YY'}$ and $c_{YY'}$,
$Y,Y'=\Lambda, \Sigma^-, \Sigma^0, \Sigma^+, \Xi^-, \Xi^0$.
In Ref.~\cite{nL,npLe}, the stiffest interaction proposed in Ref. ~\cite{BG97}, BGI, 
has been used. It is characterized by the values 
$\gamma=2$,
$a_{NN}$= - 784 MeV fm$^3$, $b_{NN}$=214.2 MeV fm$^3$,
 $c_{NN}$=1936 MeV fm$^{3 \delta}$,
$a_{\Lambda N}$= - 340 MeV fm$^3$, $c_{\Lambda N}$=1087.5 MeV fm$^{3 \gamma}$,
$a_{\Sigma N}$= - 340 MeV fm$^3$, $b_{\Sigma N}$=214.2 MeV fm$^{3}$,
$c_{\Sigma N}$=1087.5 MeV fm$^{3 \gamma}$,
$a_{\Xi N}$= - 291.5 MeV fm$^3$, $b_{\Xi N}$=0,
$c_{\Xi N}$=932.5 MeV fm$^{3 \gamma}$,
$a_{YY}$= - 486.2 MeV fm$^3$, $b_{\Lambda Y}$=0, $b_{\Xi Y}$=0,
$b_{\Sigma \Sigma}$=428.4 MeV fm$^{3}$, $c_{YY}$=1553.6 MeV fm$^{3 \gamma}$
and leads to the following values of the different interaction potential 
depths in symmetric matter at normal nuclear saturation density:
$U_{\Lambda,\Sigma}^{(N)}(n_0)=-26.6$ MeV, $U_{\Xi}^{(N)}(n_0)=-22.8$ MeV,
$U_Y^{(Y)}(n_0)$=-19.4 MeV, $U_Y^{(Y')}(n_0)$= - 38 MeV.
For the reference value $n_0/5$ BGI provides 
$U_Y^{(Y)}(n_0/5)$=-12.8 MeV, $U_Y^{(Y')}(n_0/5)$= -13.6 MeV, meaning that
it is too attractive than present double hypernuclei data indicate.

We remind that out of parameter sets proposed in Ref.~\cite{BG97}, 
BGI produces the highest neutron star maximum mass:
the maximum mass exceeds $2\,M_{\odot}$ if only
$\Lambda$'s are considered. However, if the full octet is accounted for, the
maximum neutron star mass becomes too low, possiby due to a not sufficiently
repulsive $Y$-$Y$ interaction at high densities. 

Though, as one may notice by comparison with experimental data, the $\Lambda
\Lambda$, and, probably, $\Sigma N$ interactions are too attractive at low
densitites.  The same is true for the other two potentials proposed in
Ref.~\cite{BG97}, too.  It is interesting to notice that BG-like
parameterizations in agreement with hypernuclei experimental data and able to
reach two solar mass neutron stars have been proposed in
Ref. \cite{micaela_prc85}.  For the sake of easier comparison, to see how
the phase diagram evaluated in \cite{nL,npLe} evolves when the whole baryonic
octet is accounted for, we will however stick to the original BGI in this
paper. 

To see to what extent the existence of a strangeness driven phase transition
is conditioned by the poorly-constrained $Y$-$Y$-interaction, in
section~\ref{section:modeldep} we will return to the simple case of a
$(N,\Lambda)$-mixture.  We will compare the Balberg and Gal parameterization
with an energy density functional, where the $N$-$N$ and $\Lambda$-$N$
interactions have been fitted to a microscopic Brueckner-Hartree-Fock
calculation (BSL)~\cite{BSL} and vary the parameters of the
$\Lambda$-$\Lambda$ interaction in both models. In particular we will
  show that qualitatively, in large regions of parameter space, the phase
  diagram does not depend on the exact parametrization employed.

\section{Thermodynamic analysis of the phase diagram}
\label{section:thermodynamics}

The phase diagram of a $\cal N$-component system is, at constant temperature, a
$\cal N$-dimensional volume.  The frontiers of the phase coexistence domain(s),
$\{n_i^{{\cal P}_j}\}; ~ i=1,...,{\cal N}; ~j=1,...,M$, are determined by the
$({\cal N}+1)(M-1)$ conditions of thermodynamic equilibrium between $M$ 
different phases,
\begin{eqnarray}
\left( \frac{\partial f}{\partial n_i} \right)_{{\cal P}_1}=
...=\left( \frac{\partial f}{\partial n_i} \right)_{{\cal P}_M}=\mu_i; 
~~ i=1,...,{\cal N}
\nonumber \\
\left(-f + \sum_i n_i \frac{\partial f}{\partial n_i} \right)_{{\cal P}_1}=...
=\left(-f + \sum_i n_i \frac{\partial f}{\partial n_i} \right)_{{\cal P}_M}=P, \nonumber \\
\label{eq:equil}
\end{eqnarray}
where
$f=e-Ts$, $s$ and $P$ stand for the free energy density, entropy density and,
respectively, pressure.  For a system to present a phase coexistence, its
mean-field solutions should be more expensive in terms of free energy than the
state mixing given by Eqs.~(\ref{eq:equil}).  Mathematically, this is
equivalent to the presence of a convexity anomaly of the thermodynamic
potential in the density hyperspace.  
The number of coexisting phases is determined by the
number of order parameters or, in terms of local properties, the number of
spinodal instability directions.  The last quantity is equal to the number of
negative eigenvalues ${\cal N}_{\mathit{neg}}$ of the free energy curvature matrix
$C_{ij}=\partial^2 f/\partial n_i\partial n_j$, such that $M ={\cal N}_{\mathit{neg}}+1$.

The problem of phase coexistence in a $\cal N$-component system can by 
reduced to a
problem of phase coexistence in a one-component system by Legendre
transforming the thermodynamical potential $f$ with respect to the remaining
$({\cal N}-1)$-chemical potentials~\cite{ducoin_2006}.

Under the condition of equilibrium with respect to the strong interaction,
baryonic matter is a three-component system with the densities $(n_B,n_Q,n_S)$
as degrees of freedom. It is important to remark here that the use of $S$ as a
good quantum number does not imply that strangeness density $n_S$ is
conserved. In particular, along the strangeness equilibrium trajectory
$\mu_S=0$ considered in this study $n_S$ obviously varies.

This reduces the dimensionality of the phase space from 8(=number of species)
to 3.  
To further reduce the dimensionality for studying phase coexistence,
one may then perform the Legendre transformation with respect to any set
$(\mu_B,\mu_S)$, $(\mu_S,\mu_Q)$ and $(\mu_B,\mu_Q)$.  
Formally the three choices 
should be considered in order to investigate all possible 
phase separation directions, as required by a complete study.
Within the simpler $(n, p, \Lambda)$ system studied in Ref.~\cite{npLe},
which has the same dimensionality as the full octet, we found
that the order parameter is always one dimensional.  
This means that a single Legendre transformation is enough to spot 
the thermodynamics provided that the order parameter is not orthogonal
to the controlled density.
The most convenient framework to easily access the physical trajectories 
is the one controlling the $n_B$-density:
\begin{equation}
\bar f(n_B,\mu_S,\mu_Q)=f(n_B,n_S,n_Q)-\mu_S n_S - \mu_Q n_Q.
\end{equation}
The coexisting phases, if any, will then be characterized by equal
values of $\mu_B=\partial \bar f/\partial n_B$ and $P$ and the phase
instability regions will be characterized by a back-bending behavior
of $\mu_B(n_B)|_{\mu_S,\mu_Q}$.

\section{The phase diagram of the $(n,p,Y)$-system}
\label{section:uncharge}

\begin{figure}
\begin{center}
\includegraphics[angle=0, width=0.99\columnwidth]{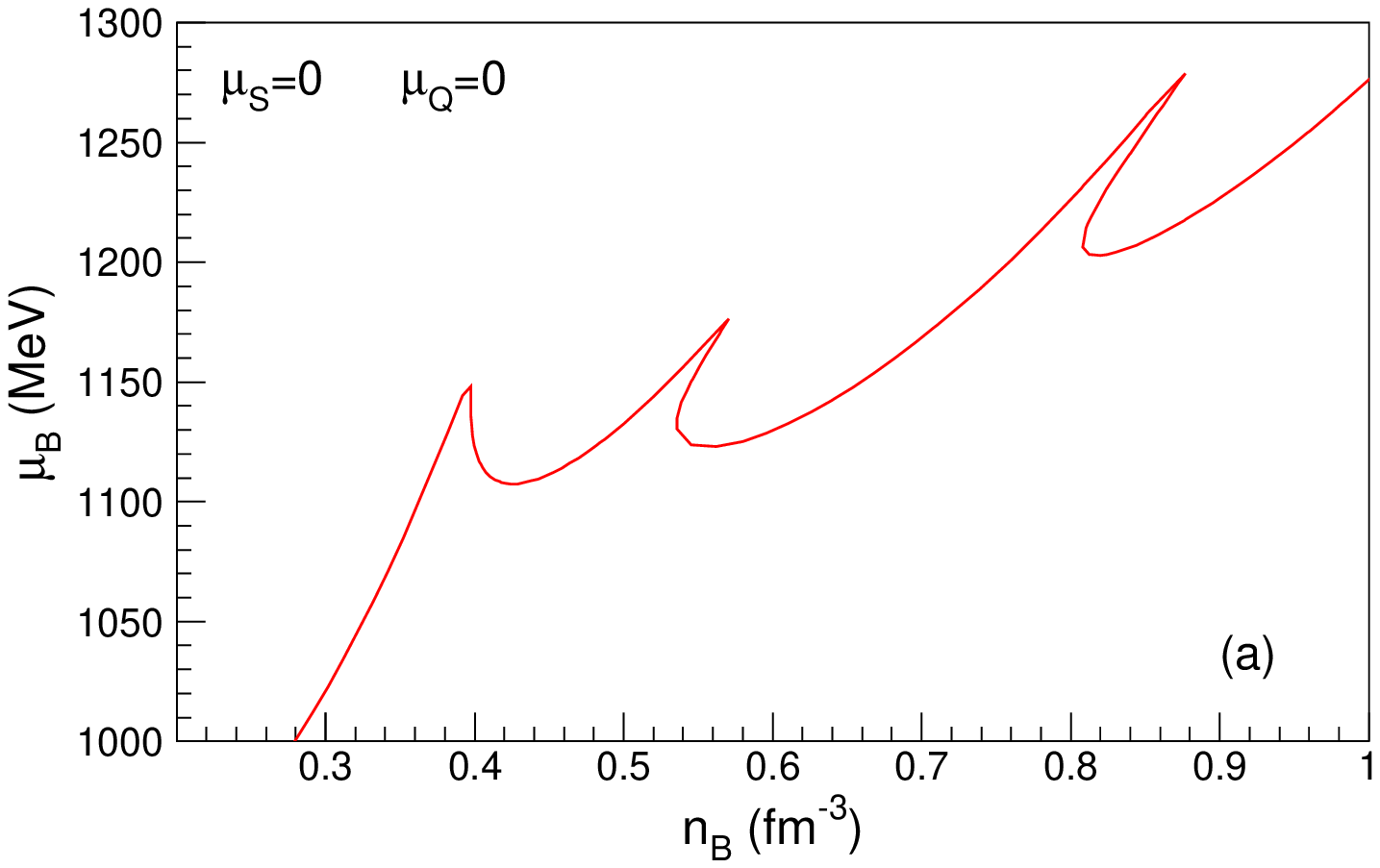}
\includegraphics[angle=0, width=0.99\columnwidth]{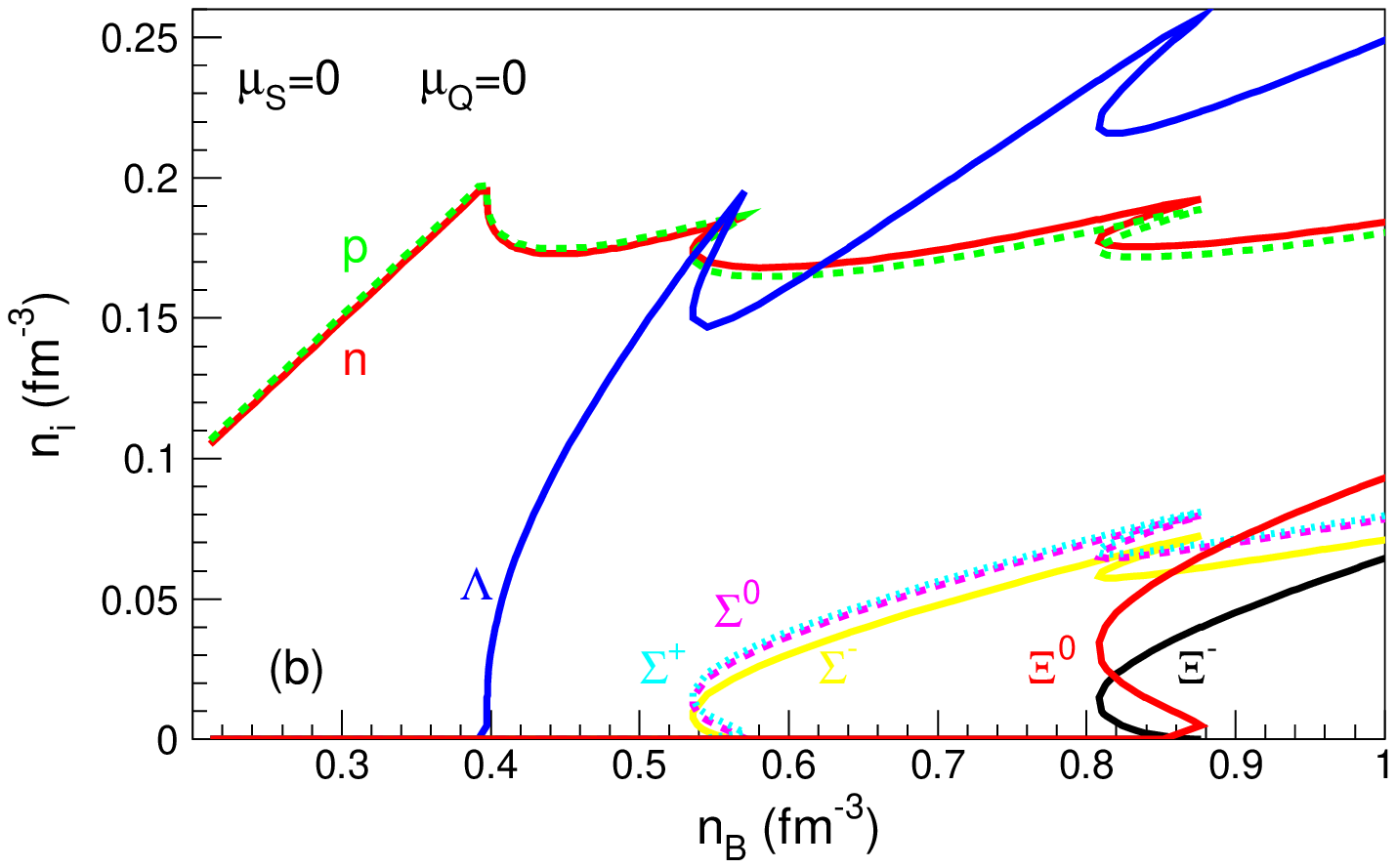}
\end{center}
\caption{(Color online)
Baryonic chemical potential (top) and particle abundances (bottom) 
as a function of baryonic density for $\mu_S=0$ and $\mu_Q=0$ at $T$=1 MeV,
employing the BGI parameterization~\cite{BG97}.
}
\label{fig:mu-rho}
\end{figure}

Within this section we will analyze the phase diagram of pure baryonic matter with
strangeness following the lines exposed in the previous section. We employ the
Balberg and Gal energy density~\cite{BG97}, parameterization BGI, see
Section~\ref{section:themodel}.  The upper panel of Fig.~\ref{fig:mu-rho}
illustrates the evolution of the baryonic chemical potential as a function of
baryonic density at constant values of $\mu_S=0$, $\mu_Q=0$ and $T$=1
MeV~\footnote{This temperature value has been chosen for computational
  convenience.  The presented results are very close to the zero temperature
  limit.} while the bottom panel depicts the abundances of different nucleonic
and hyperonic species.  Three back-bending regions in $\mu_B(n_B)$ exist.  We
can see that each back-bending is strictly correlated with the onset of new
species, and a decrease of abundances of species already present.  Upon
choosing $\mu_S=0$, $\mu_Q=0$, the hyperonic production thresholds are
exclusively determined by the particle's rest mass and the interaction
potentials. Since within the BGI parameterization, the $Y$-$Y$ and $Y$-$Y'$
interaction depends only very weakly on the particular channel, the rest mass
effects dominate. The first strange particle to appear, at about $2.6 n_0$ is
therefore the less massive one, $\Lambda$. The three quasi-degenerate
$\Sigma$-particles whose masses are 74 MeV higher than the $\Lambda$-mass, are
produced starting from about $3.6 n_0$.  The most massive hyperons, the
$\Xi$-particles, are the last to be created, at about $5.5 n_0$.  At high
densities hyperons become more abundant than nucleons. This shows that having
accurate $Y$-$Y$, $Y$-$Y'$, and $N$-$Y$ interactions is as important as having
reliable nucleonic ones.  
 
Investigation of $P(\mu_B)$ and $\bar f(n_B)$ (not shown) confirms that any
back-bending can be cured by a Maxwell construction and that the linear
combination of stable phases has a lower value for $\bar f$ than the
mean-field solution, and corresponds thus to the energetically favored
solution.  This means that 
three distinct phase coexistence regions exist, induced by the onset of each
hyperonic family.

\begin{figure}
\begin{center}
\includegraphics[angle=0, width=0.99\columnwidth]{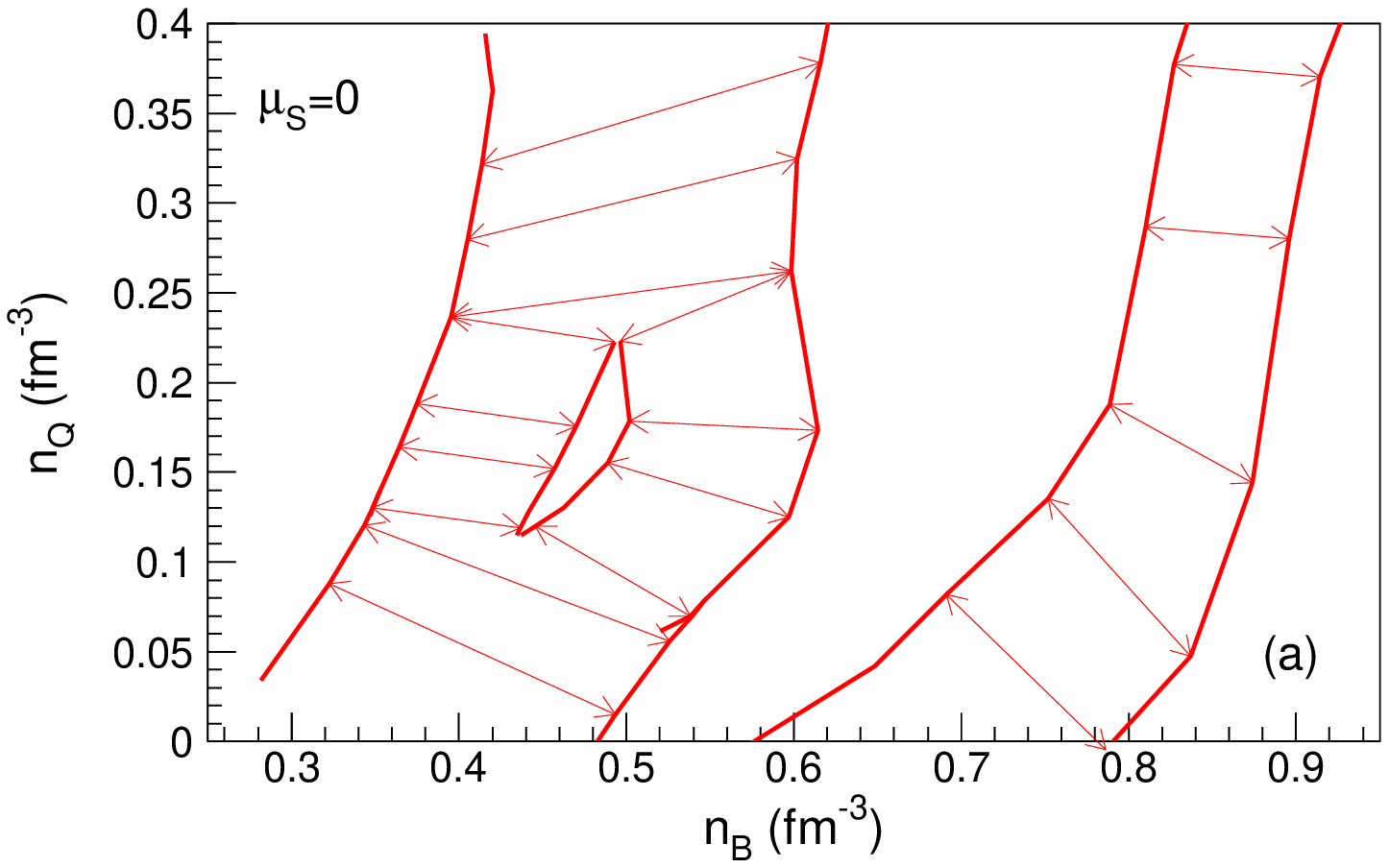}
\includegraphics[angle=0, width=0.99\columnwidth]{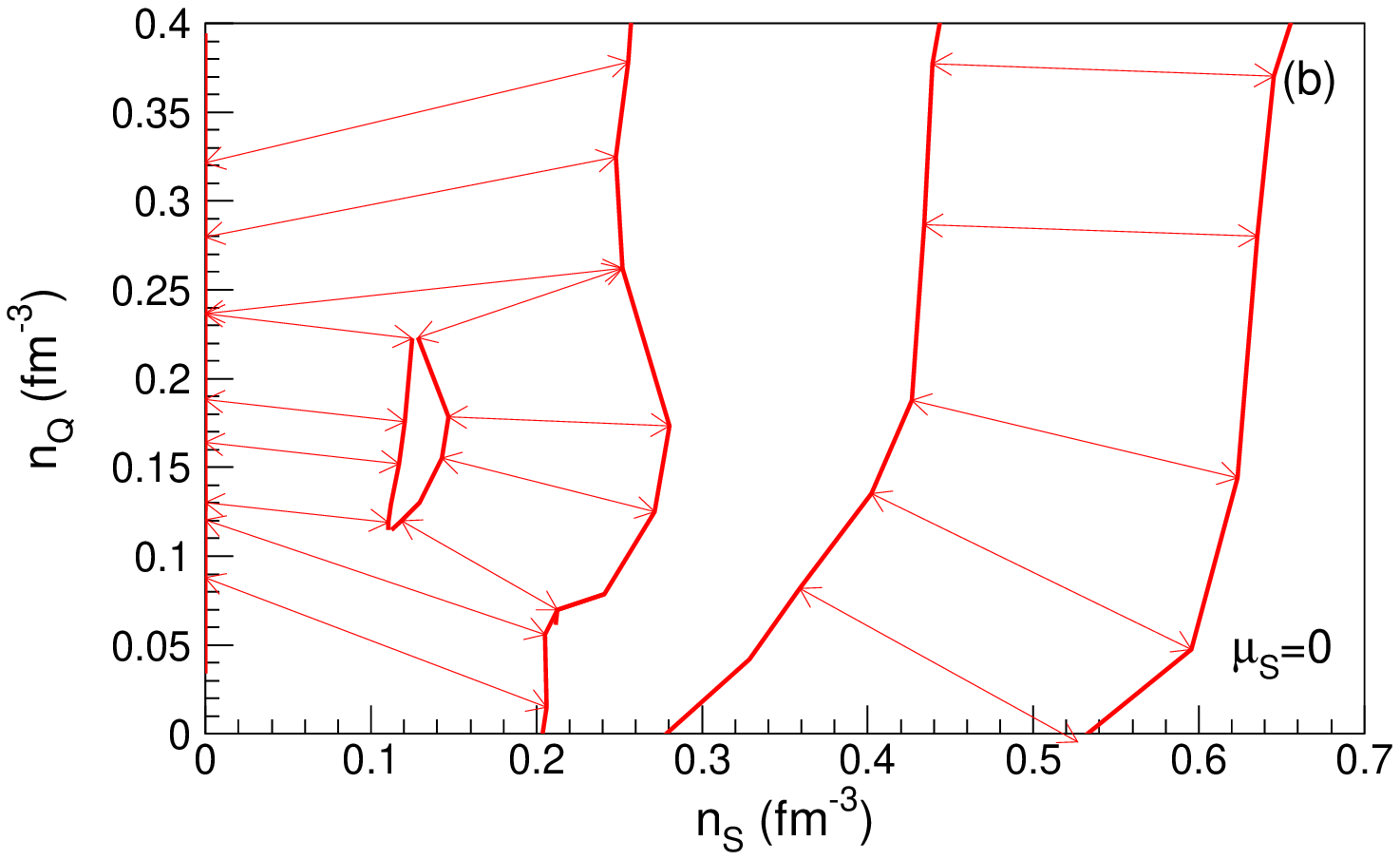}
\end{center}
\caption{(Color online)
Phase diagram of the 
$(n, p, Y)$-system 
under strangeness equilibrium  
at $T$=1 MeV as provided by the BGI parameterization~\cite{BG97}
in the $n_B$-$n_Q$ (a) and $n_S$-$n_Q$ (b) planes.
The arrows indicate the directions of phase separation.}
\label{fig:phd}
\end{figure}

Different thermodynamical conditions, i.e. different values of $(\mu_S,\mu_Q)$
and $T$, will obviously change particle production thresholds, abundances, and
the location of phase coexistence regions.  By correspondingly changing the
values of $(\mu_S,\mu_Q)$, the whole 3-dimensional phase diagram for a given
temperature can be explored. Considering that in most astrophysically
relevant situations the system is in equilibrium with respect to weak
strangeness changing interactions, the most physically relevant part of the
phase diagram is the cut corresponding to $\mu_S=0$, which will be the only
one considered within this work. The projections of the phase diagram of the
$(n,p,Y)$ mixture at the arbitrary temperature of 1 MeV to the $n_B$-$n_Q$
(panel (a)) and the $n_S$-$n_Q$-plane (panel (b)) are represented in
Fig.~\ref{fig:phd}.  The arrows indicate the directions of phase separation.
Roughly speaking, two large phase coexistence domains exist: the first one
corresponds to the appearance of $\Lambda$- and $\Sigma$-hyperons, while the
second one is due to Cascades. They are well separated and extend over a
significant total baryonic density range.

At moderate values of $\mu_Q$, where particle production is
mainly dictated by the rest mass, the thresholds for $\Lambda$- and
$\Sigma$-hyperons are pulled apart, and the phase coexistence regions
corresponding to the their respective onsets actually split up, as previously
observed for $\mu_Q = 0$, see Fig.~\ref{fig:mu-rho}.  
At more important and negative $\mu_Q$-values negatively charged particles are
favored and consequently the $\Sigma^-$-threshold is shifted to lower
densities and that for $\Sigma^+$ to higher ones. Upon increasing the absolute
value of $\mu_Q$ finally the phase coexistence region triggered by the onset
of $\Sigma$-hyperons merges with that for $\Lambda$-hyperons. The same happens
for positive values of $\mu_Q$, but with the roles of $\Sigma^-$ and
$\Sigma^+$ interchanged.

The direction of the order parameter is not constant over the phase
coexistence region.  The phase transition induced by
$\Lambda$-hyperons is always characterized by a very small component
of the order parameter along $n_Q$, as the transition is mainly
triggered by neutral $\Lambda$-hyperons, as already emphasized in
Ref.~\cite{npLe}.  The $\Sigma$-induced phase transition has a small
component along $n_Q$ when the global $\Sigma$-charge is small, that
is at low $\mu_Q$-values, and a significant component at high
$\mu_Q$-values, i.e. for a high total $\Sigma$-charge.  The
$\Xi$-induced phase transition has an order parameter with important
contribution along $n_Q$ whenever both $\Xi^0$ and $\Xi^-$ are created
as their total charge can not vanish.  At high-$\mu_Q$-values the
$\Xi^-$ production threshold is beyond the density domain considered
for this study such that only $\Xi^0$ exist and consequently the
charge dependence of the order parameter becomes very weak.

\section{The phase diagram of $(n,p,Y, e)$-system}
\label{section:neutral}

The phenomenology of baryonic matter, as the one considered above, is
purely academic.  What is pertinent from the physical point of view is
the phenomenology of electrically neutral matter, where the baryonic
charge is compensated by leptonic charge. The net charge neutrality is a
pre-requisite condition for the thermodynamic limit to exist and
corresponds to matter that constitutes compact objects where baryons
exist together with leptons and photons. It is commonly accepted that
the different sectors are in thermal and chemical equilibrium with
respect to strong and electromagnetic interactions. Chemical
equilibrium with respect to weak interactions can be satisfied or not
depending on how fast the considered astrophysical system evolves
compared with weak interaction rates.  As such, $\beta$-equilibrium is
reached in neutron stars while core-collapsing supernovae typically
evolve out of $\beta$-equilibrium.  To be as general as possible for
the moment we shall not assume $\beta$-equilibrium. As mentioned
before, we will, however, assume equilibrium with respect to
strangeness changing weak interactions.

In the mean-field approximation, the total thermodynamic potential can
be written as the sum of a baryonic, leptonic and photonic
contribution, $f=f_B+f_L+f_{\gamma}$.  Leptons and photons are well
described by fermionic and, respectively, bosonic ideal
gases~\cite{LS_npa91}.  The introduction of leptonic degrees of
freedom does not increase the dimensionality of the
problem~\cite{camille_electrons} because the strict charge neutrality
condition $n_Q=n_L$, imposed by thermodynamics, fixes $n_Q$ in terms
of leptonic density. Thus the charge degree of freedom is removed and
the associated chemical potential, $\mu_Q$ becomes ill-defined.
Within this work, $n_L = n_{e^-} - n_{e^{+}}$. The effect of other
leptons, in particular muons, is considered beyond the scope of the
present work and disregarded.  Technically, the only modification with
respect to the analysis in the case of pure baryonic matter discussed
in the previous section is the replacement of the charge density with
the (electron) leptonic one.

Adding an ideal gas contribution to the free energy might change the
convexity, i.e. the stability of the system.  Indeed, the thermodynamics of
charge neutral matter can deeply differ from that of pure baryonic matter.  As
an example, the liquid-gas (LG) phase transition taking place in nuclear
matter at sub-saturation densities is strongly
quenched~\cite{camille_electrons,providencia_electrons} by the presence of
electrons via the charge neutrality condition. This is due to the fact that a
first order (LG) transition is associated with a macroscopic density
fluctuation in direction of the order parameter. In the case of the nuclear LG
transition, the order parameter has a strong component in charge direction,
implying a macroscopic charge density fluctuation. This fluctuation is,
however, strongly suppressed by the high incompressibility of the electron
gas.

Ref.~\cite{npLe} shows that, for the $(n,p,\Lambda)$ system, the
strangeness-driven phase transition is essentially not affected by the
electrons.  This is not surprising because $\Lambda$ fluctuations are
poorly correlated to the electric charge, see previous section, too.
The situation is different here, because of the presence of charged
hyperons. As discussed before, we can see in Fig.~\ref{fig:phd} that
the order parameter has a significant component
along the charge density especially for the $\Xi$-induced transition
at low $\mu_Q$-values, and it is in this domain that we expect the
most dramatic alteration of the phase coexistence region.

\begin{figure}
\begin{center}
\includegraphics[angle=0, width=0.99\columnwidth]{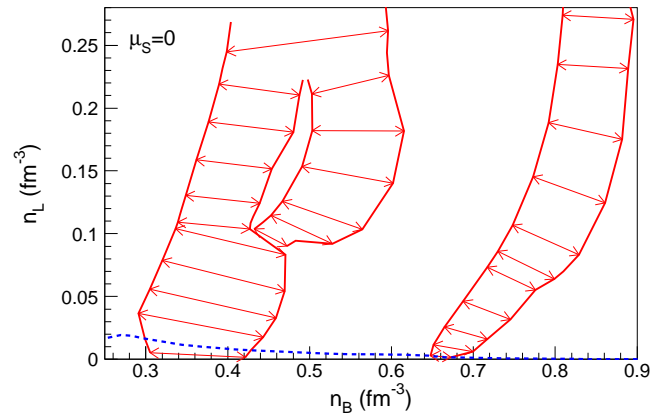}
\end{center}
\caption{(Color online)
Phase diagram of the $(n,p,Y, e)$-system
under strangeness equilibrium at $T$=1 MeV
as provided by BGI parameterization \cite{BG97} in the $n_B$-$n_L$-plane.
The dotted curve marks the path corresponding to $\beta$-equilibrium.
}
\label{fig:phd_+e}
\end{figure}

The phase diagram of the $(n,p,Y, e)$-system under strangeness
equilibrium at $T$=1 MeV is displayed in Fig.~\ref{fig:phd_+e} in the
plane $n_B$-$n_L$.  As before, the arrows mark the directions of the
order parameter.  The pattern of the phase diagram is roughly the same
as for pure baryonic matter: depending on $\mu_L$, $\Lambda$- and
$\Sigma$-hyperons are responsible for one or two phase transitions
which extend over $0.3 \lesssim n_B \lesssim 0.6$ fm$^{-3}$ and a
$\Xi$-induced phase transition occurs at higher baryonic densities.
The most important shrinking of the phase coexistence is obtained at
low $n_L$-values.  
The direction of phase separation is rotated in the sense that its
component along $n_L$ gets smaller, which is expected since large
electron density fluctuations are effectively suppressed.

\section{Model and parameter dependence}
\label{section:modeldep}

The predictions of a phenomenological density-functional model depend
dramatically on the functional form assumed for the energy
density and the employed values of the coupling constants.  As
discussed in Section~\ref{section:themodel}, the functional form of
the energy density in a non-relativistic phenomenological model is
subject to large arbitrariness.  The same is true for the coupling
constants as the experimental data (a) correspond exclusively to low
matter density, (b) are insufficient to constrain all the parameters
of the potential energy functional and (c) are often subject to large
uncertainties, especially for the $Y$-$Y(Y')$ channels.  As a
consequence, instead of one particular functional with one parameter
set, one should rather consider different parameter sets and
functional forms, satisfying the experimental conditions.

For this reason, we will first examine the correlation between the
existence of the phase transition and the parameters of the Skyrme-based
BG~\cite{BG97} energy density functional.  To avoid proliferation
of unconstrained parameters, the issue is considered in the simple
case of a $(n,p,\Lambda)$ mixture, which nevertheless satisfies the
basic requirement of accounting for all relevant degrees of freedom,
$B$, $S$, and $Q$ 
 The two interaction channels
which can be responsible of the phase transition are the $N$-$\Lambda$
and the $\Lambda$-$\Lambda$ one. Since the $Y$-$Y$ interactions are
poorly known, we first consider the extreme situation where the
$\Lambda$-$\Lambda$ coupling is completely absent.

\subsection{The $N$-$Y$-interaction}

The parameters of the $N$-$\Lambda$-channel, $\gamma$, $a_{\Lambda N}$
and $c_{\Lambda N}$, are considered as free variables which have to
satisfy the unique condition $U_{\Lambda}^{(N)}(n_0)=-26.6$ MeV,
keeping for simplicity the reference value of BGI. We consider $1.1
\leq \gamma \leq 3$, $-1000$ MeV fm$^3 \leq a_{\Lambda N} \leq -100$
MeV fm$^3$ and, in each case, calculate
$c_{YN}=(U_{\Lambda}^{(N)}(n_0)-a_{N \Lambda}\cdot
n_0)/n_0^{\gamma}$. The nuclear part remains the same as for BGI.
\begin{figure}
\begin{center}
\includegraphics[angle=0, width=0.9\columnwidth]{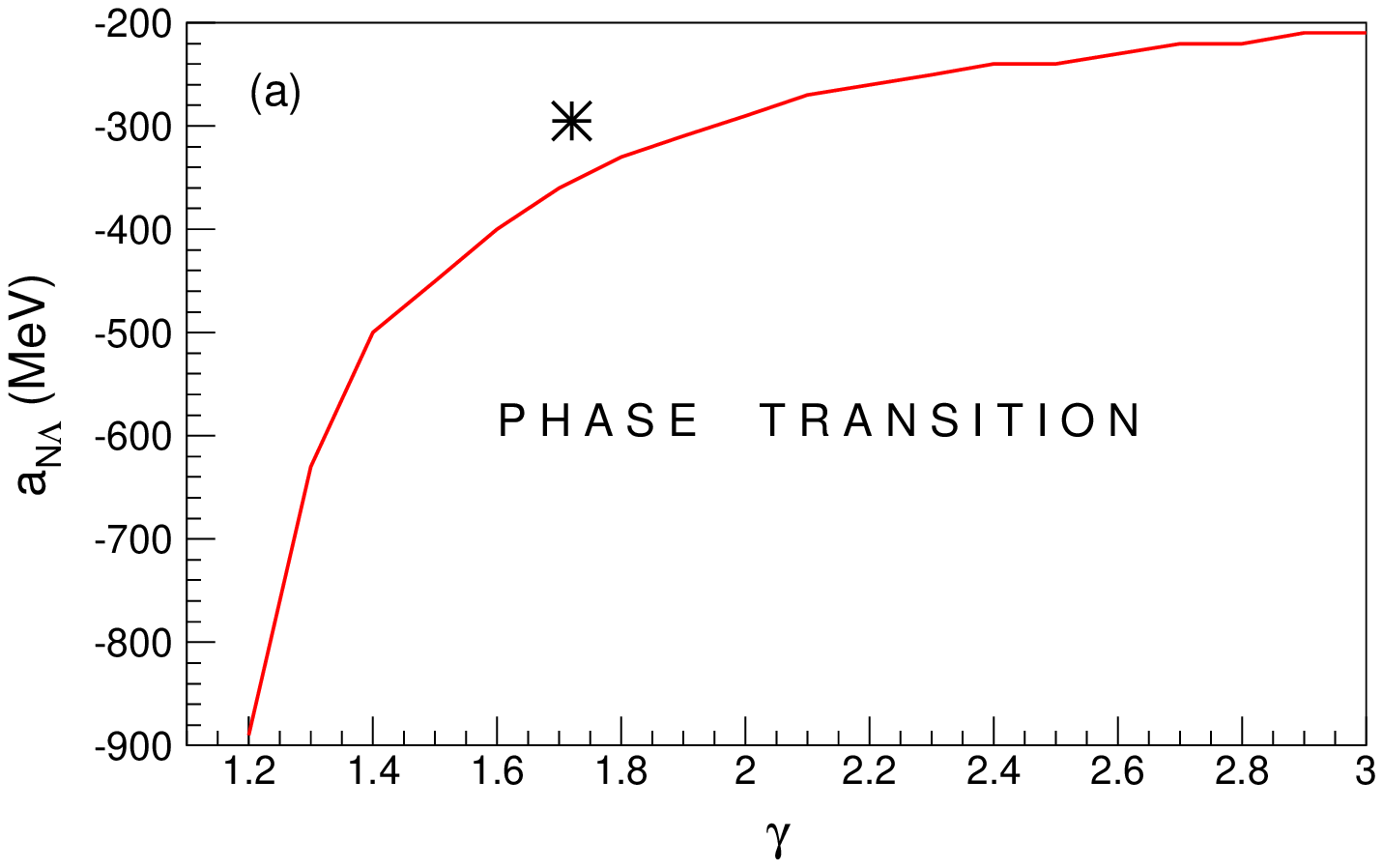}
\includegraphics[angle=0, width=0.9\columnwidth]{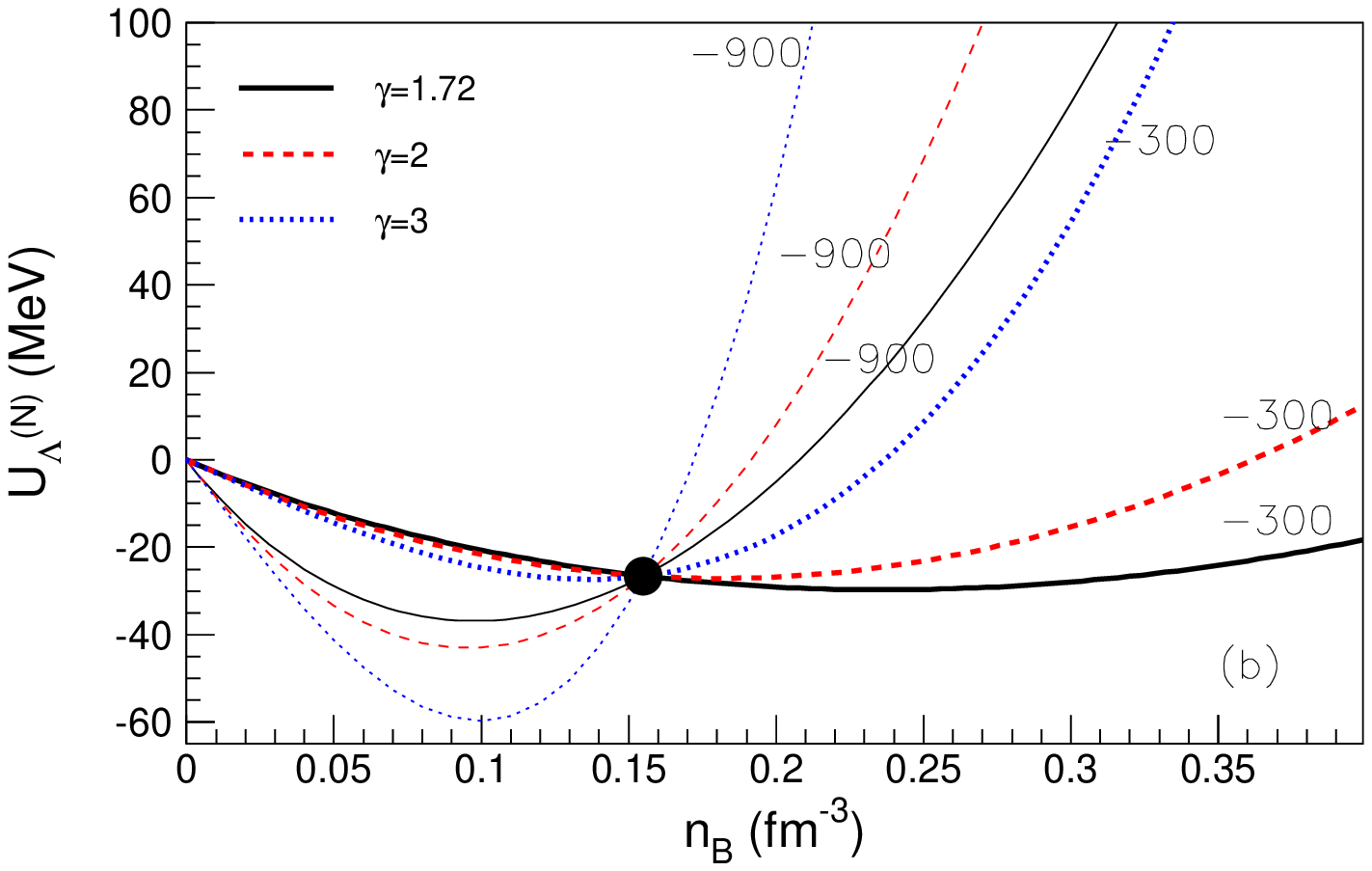}
\includegraphics[angle=0, width=0.9\columnwidth]{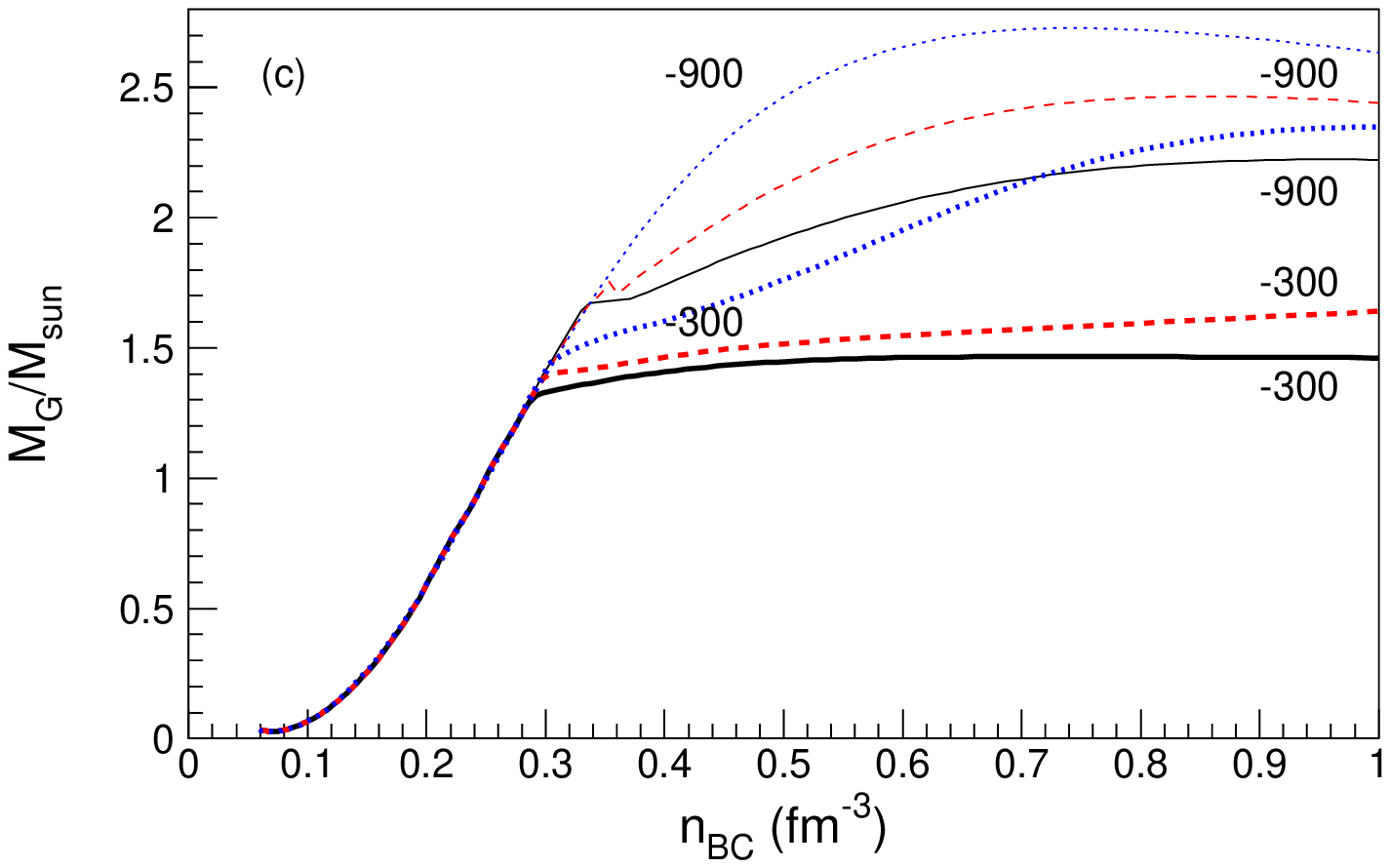}
\includegraphics[angle=0, width=0.9\columnwidth]{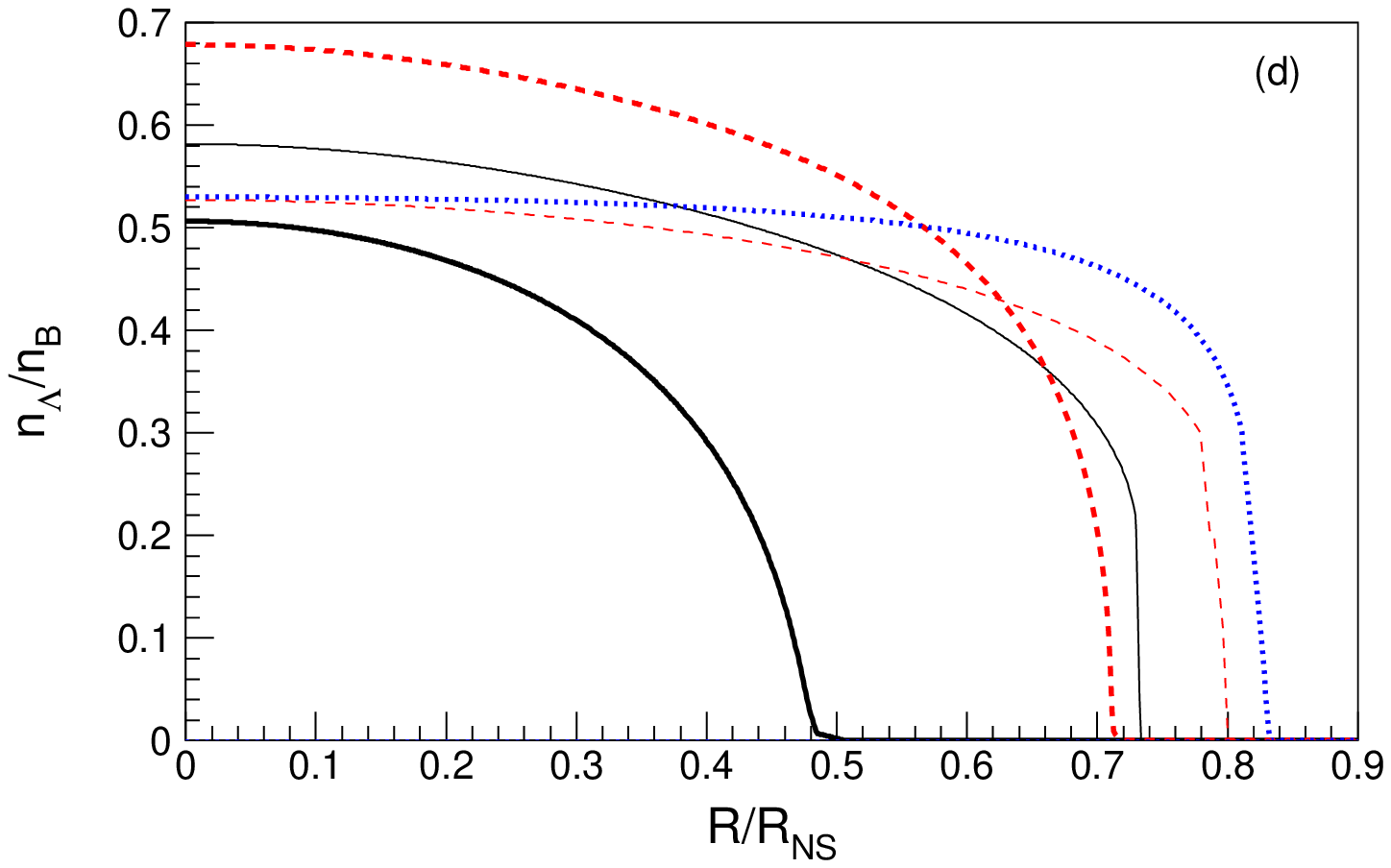}
\end{center}
\caption{(Color online) BG parameterization without
  $\Lambda$-$\Lambda$-interaction: (a) Limiting values of the coupling
  constant $a_{N \Lambda}$ for which, at different $\gamma$, the symmetric
  $(N,\Lambda)$-system at $T$=0 manifests strangeness driven phase transition
  along $\mu_S=0$. 
  The star marks the $(\gamma,a_{N \Lambda})$ values corresponding to the BSL
  $NY$-interaction (see text). 
  (b) Nucleonic density
  dependence of the $\Lambda$ potential in uniform symmetric nuclear matter
  $U_{\Lambda}^{(N)}(n_p + n_n)$ for different $(\gamma$,$a_{YN}$[MeV fm$^3])$
  sets: (1.72,-900), (1.72,-300), (2,-900), (2,-300), (3,-900) and (3,-300).
  $c_{YN}$ is determined such as $U_{\Lambda}^{(N)}(n_0)$=-26.6 MeV~\cite{BG97},
  see text; 
  (c) Neutron star mass as a function of central baryon density for
  the $(\gamma$,$a_{YN})$ sets considered in (b);
  (d) $\Lambda$ relative abundances as a function of normalized
  distance from the star center for the maximum mass configuration
  and the $(\gamma$,$a_{YN})$ sets considered in (b).}
\label{fig:instab_BG_noYY}
\end{figure}

The upper panel of Fig. \ref{fig:instab_BG_noYY} plots, as a function
of the stiffness parameter $\gamma$, the maximum values of the
coupling constant $a_{N \Lambda}$ for which symmetric
$(N,\Lambda)$-matter manifests phase coexistence along $\mu_S=0$.  As
one may note, irrespective of $\gamma$, there is a wide range of
values for the attractive $Y$-$N$ coupling meaning that, in this model,
phase coexistence in hyperonic matter is not conditioned by the
$Y$-$Y$-interaction.  The behavior of the $\Lambda$-potential in
symmetric nuclear matter, $U_{\Lambda}^{(N)}(n_B)=\partial
e_{pot}(n_N,n_{\Lambda})/\partial n_{\Lambda}|_{n_{\Lambda}=0}$, as a
function of nucleonic density is illustrated in panel (b) of
Fig.~\ref{fig:instab_BG_noYY} for few representative $\gamma$-values
($\gamma=$1.72, 2 and 3)
and coupling constants situated at the extremities of the considered
range ($a_{N \Lambda}=$ -900 and -300 MeV fm$^3$), 
both inside and outside the domain compatible with phase coexistence, 
as indicated on the figure.  
We can see that a wide variety of density behaviors are compatible with the
presence of a phase transition.

The new very precise astrophysical measurements of neutron star masses
close to two solar masses \cite{demorest,antoniadis} represent a
validity test for any astrophysical equation of state.  As the rich
recent literature testifies, this supplementary piece of information
can neither confirm nor rule out the presence of hyperons in neutron
stars. Indeed, while it is true that in principle any extra degree of
freedom softens the EOS and, thus, lowers the maximum mass of the
star, various models 
\cite{Bednarek11,micaela_prc85,sedrakian_aa2012,weissenborn11a,weissenborn11b} 
prove that hyperons are compatible with the two solar mass constraint.  
The predictions of the $\beta$-equilibrium EOS
at zero temperature for the neutron star mass as a function of central
density obtained by solving the Tolman-Oppenheimer-Volkoff
(TOV)~\cite{TOV} equations for hydrostatic equilibrium of a spherical
star,
\begin{eqnarray}
\frac{d P(r)}{d r}&=&-\frac{G}{r^2} \left[\epsilon(r)+\frac{P(r)}{c^2} \right]
\left[M(r)+4 \pi r^3 \frac{P(r)}{c^2}\right] \nonumber \\
&\cdot& \left[ 1-\frac{2 G M(r)}{c^2 r}\right]^{-1};
\nonumber \\
\frac{d M(r)}{d r}&=&4 \pi \epsilon(r) r^2
\end{eqnarray}
are represented in panel (c) of Fig.~\ref{fig:instab_BG_noYY}
for the parameter sets considered in panel (b).

Eq. (\ref{eq:coupling}) shows that, for the presently considered functional
form of the energy density Eq.(\ref{eq:epot_BG}), a strong attraction at low
density is associated with a strong repulsion at high density through the
fixed $U_{\Lambda}^{(N)}(n_0)$.  It is this peculiarity that makes possible to
produce, by the most repulsive potentials, gravitational masses which largely
exceed the reference $2 M_{\odot}$ limit and have an important
$\Lambda$-hyperon fraction.  It is, however, important to stress that these
results have to be considered as qualitative, because of the artificial
absence of other hyperons than $\Lambda$'s. The inclusion of the full
octet will obviously influence the mass-radius relationship quantitatively.

The bottom panel of Fig. \ref{fig:instab_BG_noYY} depicts, for the above
considered $NY$ interaction potentials and the maximum mass neutron star
configuration, the $\Lambda$-relative abundances as a function of normalized
distance from the star center.  The reason why the curve corresponding to
(3,-900) is missing is the extremely repulsive potential which prevent
$\Lambda$s to appear.  For the other interaction potentials one can see that
hyperons not only exist, but they are abundant and populate most of the star's
volume.  The different central baryonic density values which correspond to the
maximum mass configuration prevent a straightforward parallelism among the
stiffness of the potential on one hand and the hyperonic relative density in
the core and its extension on the other hand. This becomes obvious
observing that the lowest and the highest fractions in the star core
correspond to the softest considered potentials.

\subsection{The $Y$-$Y$-interaction}

We now turn to study the effect of the $\Lambda$-$\Lambda$
interaction, both on the existence of the phase transition and on the
maximum NS mass.  To keep the same framework we shall consider the
original BGI parameterization for the $N$-$\Lambda$ channel, and the BG
functional dependence in the $\Lambda$-$\Lambda$ channel.  Again, as
in the case of the $N$-$\Lambda$ interaction, we vary the
$\Lambda$-$\Lambda$ parameters keeping the $\Lambda$-potential in
uniform $\Lambda$-matter at $1/5$ of nuclear saturation density fixed,
$U_{\Lambda}^{(\Lambda)}(n_0/5)$, which leads to $c_{\Lambda
  \Lambda}=\left(U_{\Lambda}^{(\Lambda)}(n_0/5)-a_{\Lambda \Lambda}
n_0/5 \right) \cdot 2/(\gamma+1)/(n_0/5)^{\gamma}$. 
The adopted value shall be $U_\Lambda^{(\Lambda)} (n_0/5)  = $ - 0.67 MeV, 
see the discussion in Section~\ref{section:themodel}.  We then consider
different parameter sets $(\gamma, a_{N \Lambda}, a_{\Lambda
  \Lambda})$ in the ranges $1.2 \leq \gamma \leq 3$, $-1000 $ MeV
fm$^3 \leq a_{N \Lambda} \leq -200$ MeV fm$^3$ and $-1000 $ MeV fm$^3
\leq a_{\Lambda \Lambda} \leq -100$ MeV fm$^3$.

\begin{figure}
\begin{center}
\includegraphics[angle=0, width=0.99\columnwidth]{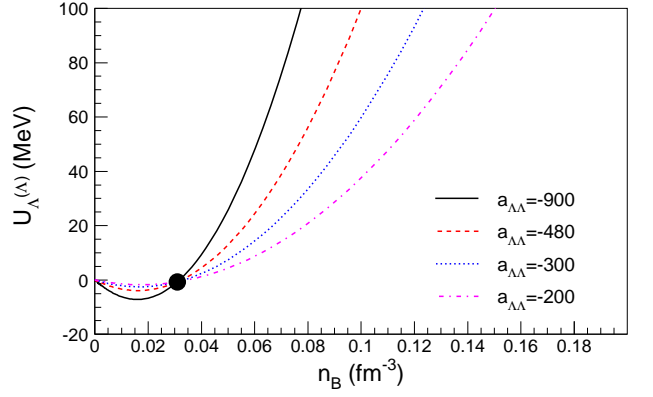}
\end{center}
\caption{(Color online) 
$\Lambda$-potential in uniform $\Lambda$-matter as a function of density
corresponding to BGI parameterization with modified
$\Lambda$-$\Lambda$ interaction: $a_{\Lambda \Lambda}$=-900, -480,
-300 and -200 MeV fm$^3$ and $\gamma = 2$.  
}
\label{fig:BGI+modifYY_ULL}
\end{figure}

\begin{figure}
\begin{center}
\includegraphics[angle=0, width=0.99\columnwidth]{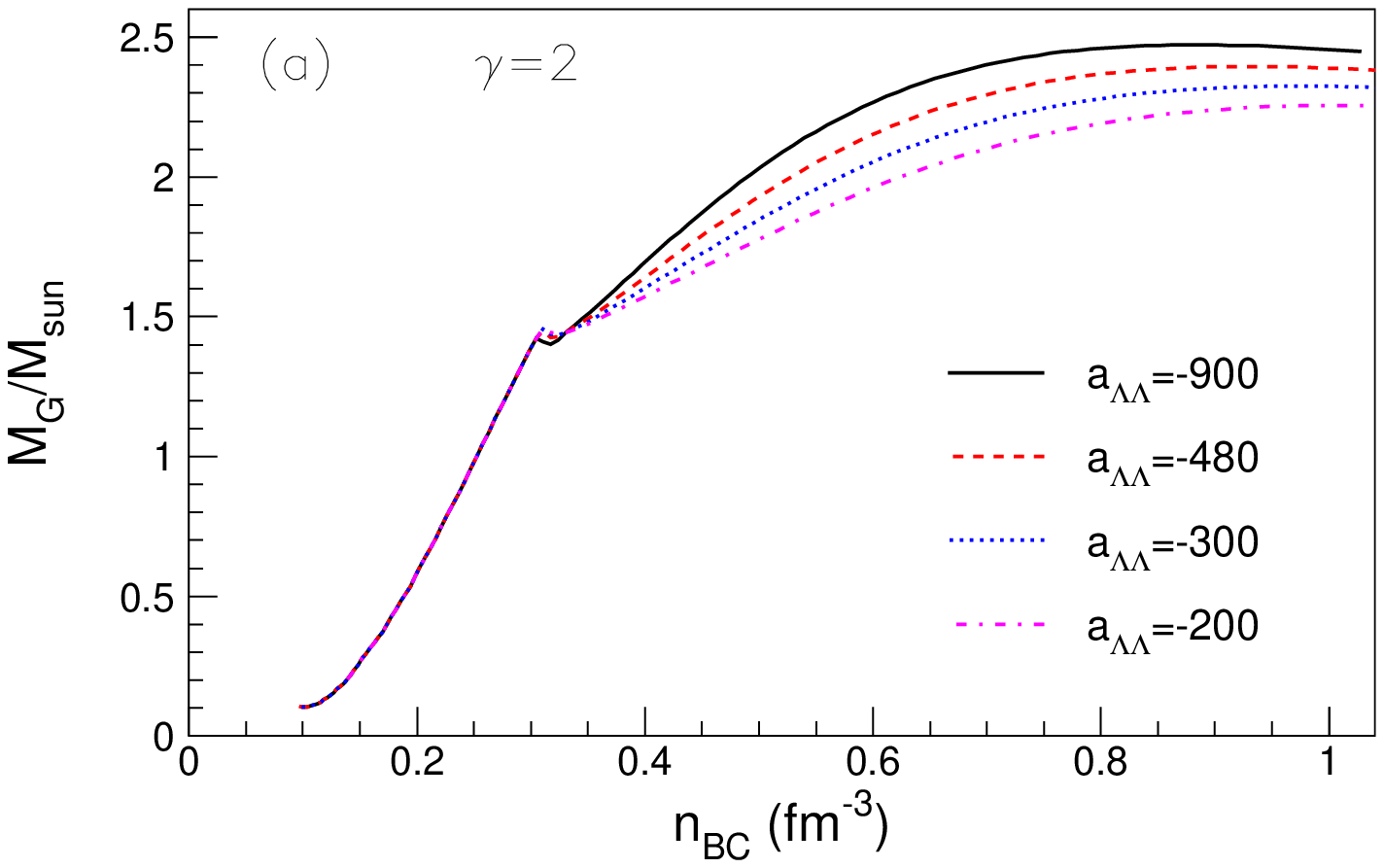}
\includegraphics[angle=0, width=0.99\columnwidth]{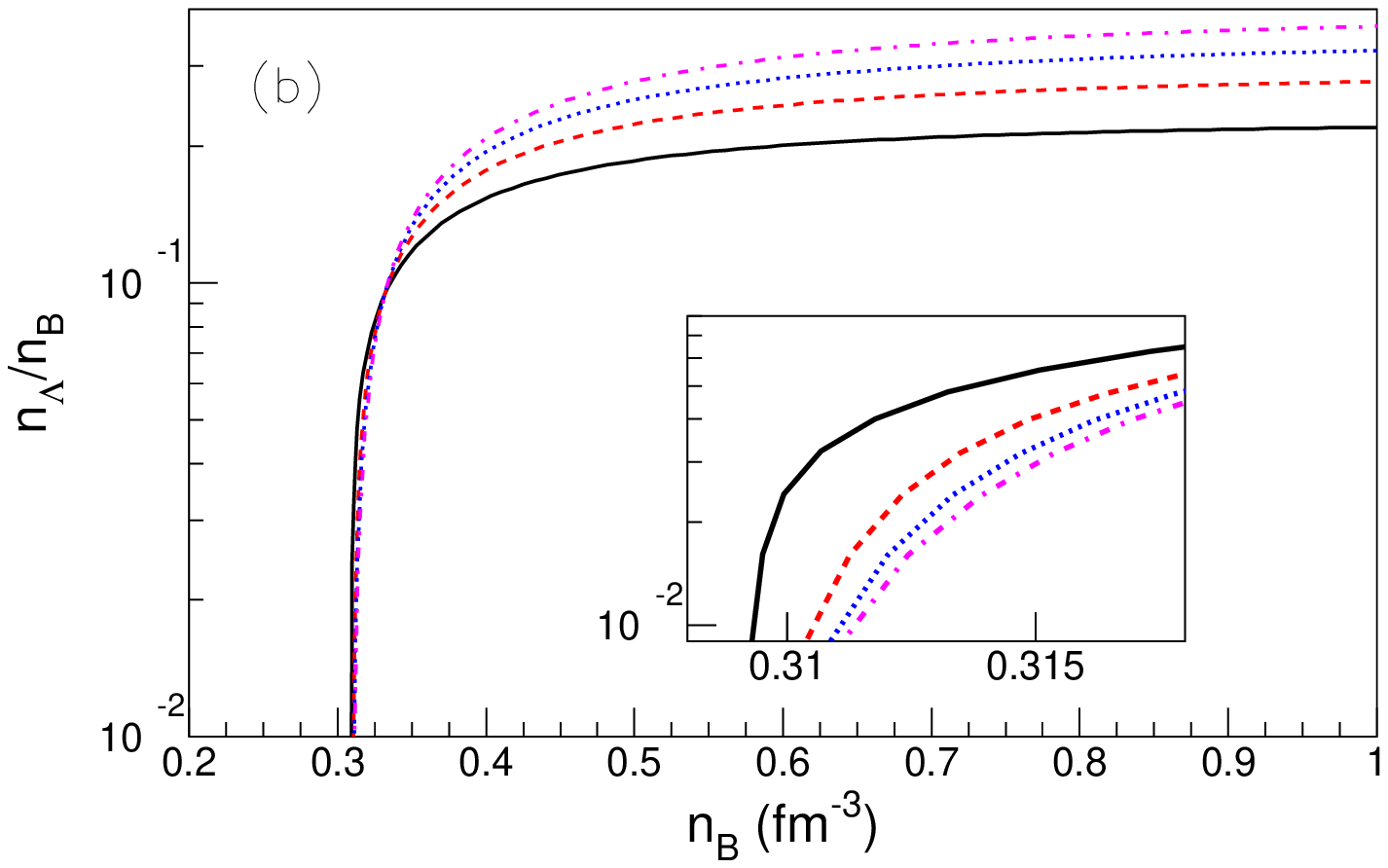}
\end{center}
\caption{(Color online) 
(a) Neutron star mass as a function of central density and 
(b) Relative $\Lambda$ abundances as a function of baryonic density along
the $\beta$-equilibrium path for the effective interaction potentials considered
in Fig. \ref{fig:BGI+modifYY_ULL}.} 
\label{fig:BGI+modif_YY_star}
\end{figure}

Adding the $\Lambda$-$\Lambda$ long-range attractive - short-range
repulsive interaction, the $(n,p,\Lambda)$ toy system manifests phase
coexistence in a much broader parameter range. 
More precisely, a
strangeness-driven phase transition along the $\mu_S=0$ path is
obtained for almost all considered sets.

Fig. \ref{fig:BGI+modifYY_ULL} illustrates
$U_{\Lambda}^{(\Lambda)} (n_{\Lambda} = n_B)$ for $\gamma=2$ and
$a_{\Lambda \Lambda}$=-900, -480, -300, -200 MeV. As in the case of the
$N$-$\Lambda$-channel, a strong attraction leads to a steep rise at
high densities and a deep minimum localized at low densities, due to
the fact that we fix the value at $n_0/5$ and the correlation between
  attraction and repulsion for the BG functional form.

Fig. \ref{fig:BGI+modif_YY_star} depicts the predictions of these potentials
for the NS gravitational mass as function of central density (panel (a))
together with the $\Lambda$-relative abundances as a function of baryonic
density along the beta-equilibrium trajectory (panel (b)).  In all cases, for
the $N$-$\Lambda$-channel the BGI parameter values have been employed.  The
relative ordering of the various curves is easily understandable for high
central densities where the short range repulsion is effective: the stronger
is the $\Lambda$-$\Lambda$ repulsion, the smaller is the relative $\Lambda$
density and the larger the obtained NS mass.  Equally predictable is the fact
that small differences in the low density attractive part of the potential
result in minor modifications of the $\Lambda$-production threshold, to a
large extend dictated by the $N$-$Y$ interaction.  Indeed, for the most
attractive considered potential, $\Lambda$-hyperons emerge at a baryonic
density only $3 \cdot 10^{-4}$ fm$^{-3}$ lower that than the one corresponding
to the least attractive potential.

The calculations presented so far were all obtained with a
phenomenological non-relativistic functional, that proposed by Balberg and
Gal~\cite{BG97}, both for the $N$-$\Lambda$ and the
$\Lambda$-$\Lambda$ channel.  One can therefore wonder if the observed
phase transition is not a pathology of the assumed and largely
arbitrary functional form of the energy density.

\subsection{BHF $N$-$Y$ interaction potentials}

For more than fifteen years different microscopically motivated $N$-$N$ and
$N$-$Y$ interaction potentials have been proposed.  These functionals have all
been adjusted to Brueckner-Hartree-Fock calculations 
hyper-nuclear matter based on different bare $N$-$N$ and $N$-$Y$ interactions,
and are designed to perform calculations of
hypernuclei~\cite{cugnon_2000,vidana_2001, zhou,schulze} and more recently
hyper-nuclear matter~\cite{BSL, SR}.  In all of these potentials, $Y$-$Y$
  interactions have been disregarded because of missing experimental
  constraints for the basic two-particle $Y$-$Y$ interaction and the
  difficulties observed with BHF calculations including the unsufficiently
  constrained bare $Y$-$Y$-interaction.

The two parametrizations designed for hyper-nuclear matter, Refs. \cite{BSL,SR},
rely on the same energy density functional,
\begin{eqnarray}
e_{N \Lambda}^{(BSL)}&=&\left(a_{\Lambda}^0+a_{\Lambda}^1 x+a_{\Lambda}^2 x^2 \right)n_N n_{\Lambda}
+\left(b_{\Lambda}^0+b_{\Lambda}^1 x+b_{\Lambda}^2 x^2 \right) n_N^{c_{\Lambda}} n_{\Lambda}  \nonumber \\
&+&a^{(BSL)}_{\Lambda \Lambda} n_N^{c^{(BSL)}_{\Lambda \Lambda}} n_{\Lambda}^{d_{\Lambda \Lambda}+1},
\label{eq:epot_burgio}
\end{eqnarray}
where $x=n_p/n_N$. 
They differ in the coupling constants values as the BHF calculations correspond to
various treatments of the three-body forces and $Y$-$N$-potentials and predict 
significantly different $\Lambda$- and $\Sigma^-$-abundances \cite{SR}.
For our application we have chosen to use the Burgio-Schulze-Li parameterization \cite{BSL}
because of its stiffer $U_{\Lambda}^{(N)}(n_N)$ dependence.

Despite the functional dissimilarity between Eqs. (\ref{eq:epot_burgio}) 
and (\ref{eq:epot_BG}), the two parameterizations can be bridged
via the $\Lambda$-potential in uniform symmetric nuclear matter,
\begin{eqnarray}
U_{\Lambda}^{(N)}(n_N)
&=&\frac{\partial e_{\mathit{pot}}^{(BSL)}(n_N,n_{\Lambda})}{\partial n_{\Lambda}}|_{n_{\Lambda}=0} \nonumber \\
&=&
\frac{\partial e_{N\Lambda}^{(BSL)}(n_N,n_{\Lambda})}{\partial n_{\Lambda}}|_{n_{\Lambda}=0} \nonumber \\
&=&
\left(a_{\Lambda}^0 +\frac{a_{\Lambda}^1}2 +\frac{a_{\Lambda}^2}4 \right) n_N+
\left(b_{\Lambda}^0 +\frac{b_{\Lambda}^1}2 +\frac{b_{\Lambda}^2}4 \right) n_N^{c_{\Lambda}}.
\nonumber \\
\label{eq:ULNschulze}
\end{eqnarray}
which, in both cases, is a polynomial in the baryonic density.
Similarly to Eq. (\ref{eq:coupling}) a correlation between 
short- and long-range interactions is present in
Eq.~(\ref{eq:ULNschulze}). 
We note that Eq. (\ref{eq:ULNschulze}) can be mapped onto  
$\partial e_{pot}^{(BG)}/\partial n_{\Lambda}|_{n_{\Lambda}=0}$ provided that
$\gamma = c_{\Lambda} = 1.72,
a_{N\Lambda} = a^0_\Lambda + a^1_\Lambda /2 + a^2_\Lambda/4 = -294.75 $MeV fm$^3,
c_{N\Lambda} = (b^0_\Lambda + b^1_\Lambda /2 + b^2_\Lambda /4) = 462.75$ 
MeV fm$^{3 \gamma}$. 
The (1.72,-294.75) point is represented in Fig. \ref{fig:instab_BG_noYY} (a) 
by a star and sits outside the phase coexistence domain of a symmetric 
$(N,\Lambda)$
mixture at strangeness equilibrium.
We note that these values are very close to ($\gamma=1.72$, 
$a_{\Lambda N}=-300$ MeV fm$^3$) for which
$U_{\Lambda}^{(N)} (n_N)$ is depicted in Fig. \ref{fig:instab_BG_noYY}.

\begin{figure}
\begin{center}
\includegraphics[angle=0, width=0.99\columnwidth]{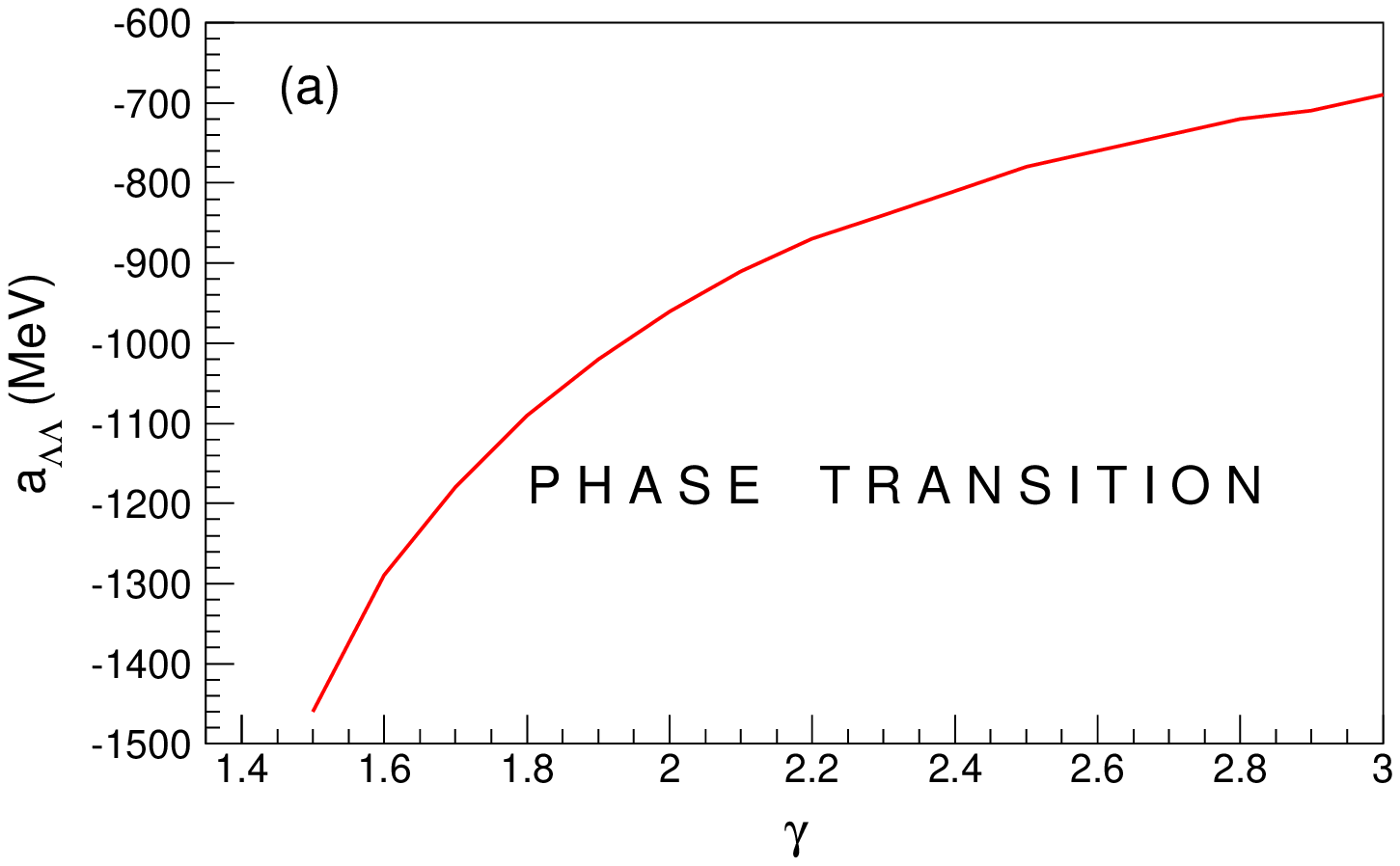}
\includegraphics[angle=0, width=0.99\columnwidth]{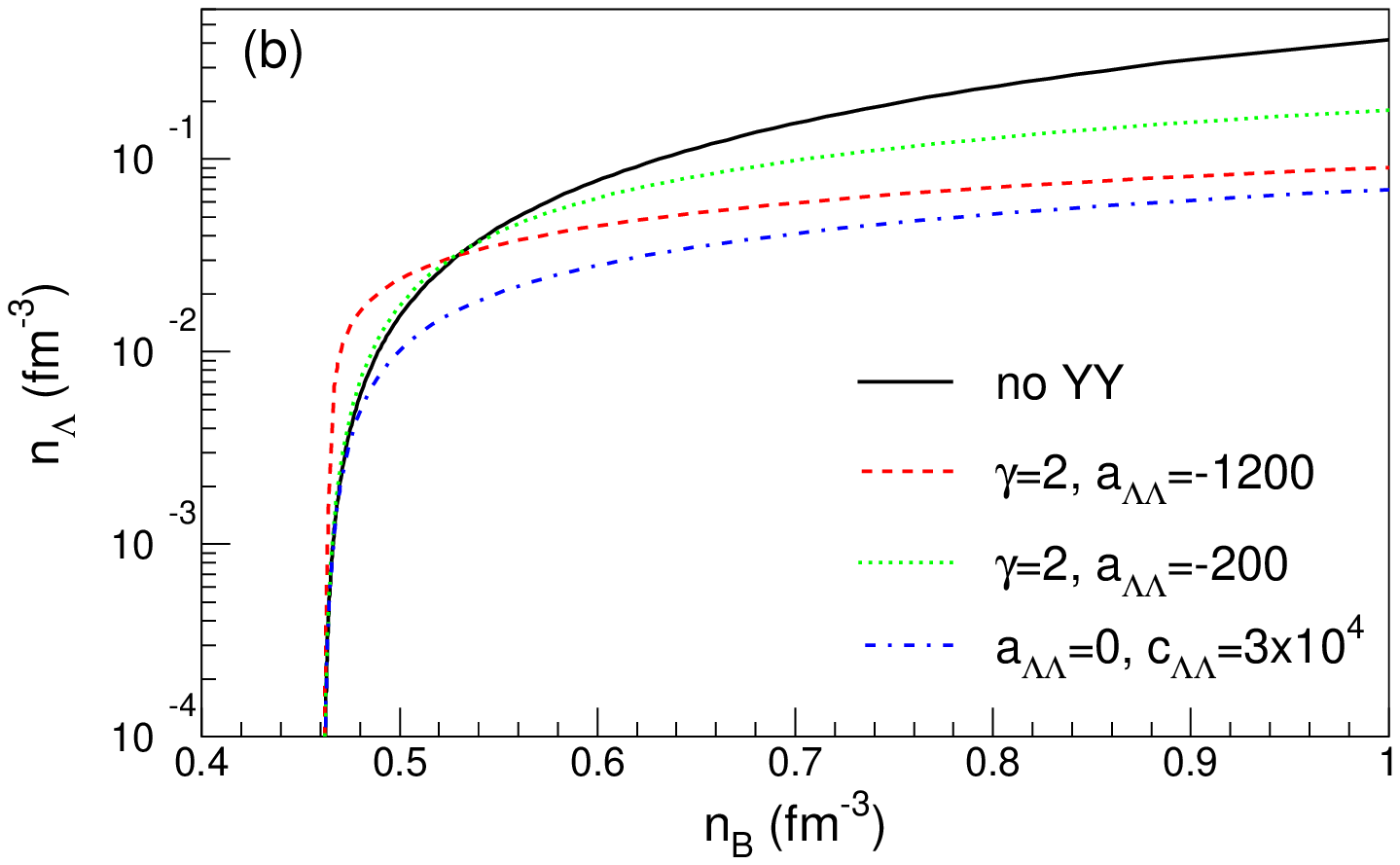}
\includegraphics[angle=0, width=0.99\columnwidth]{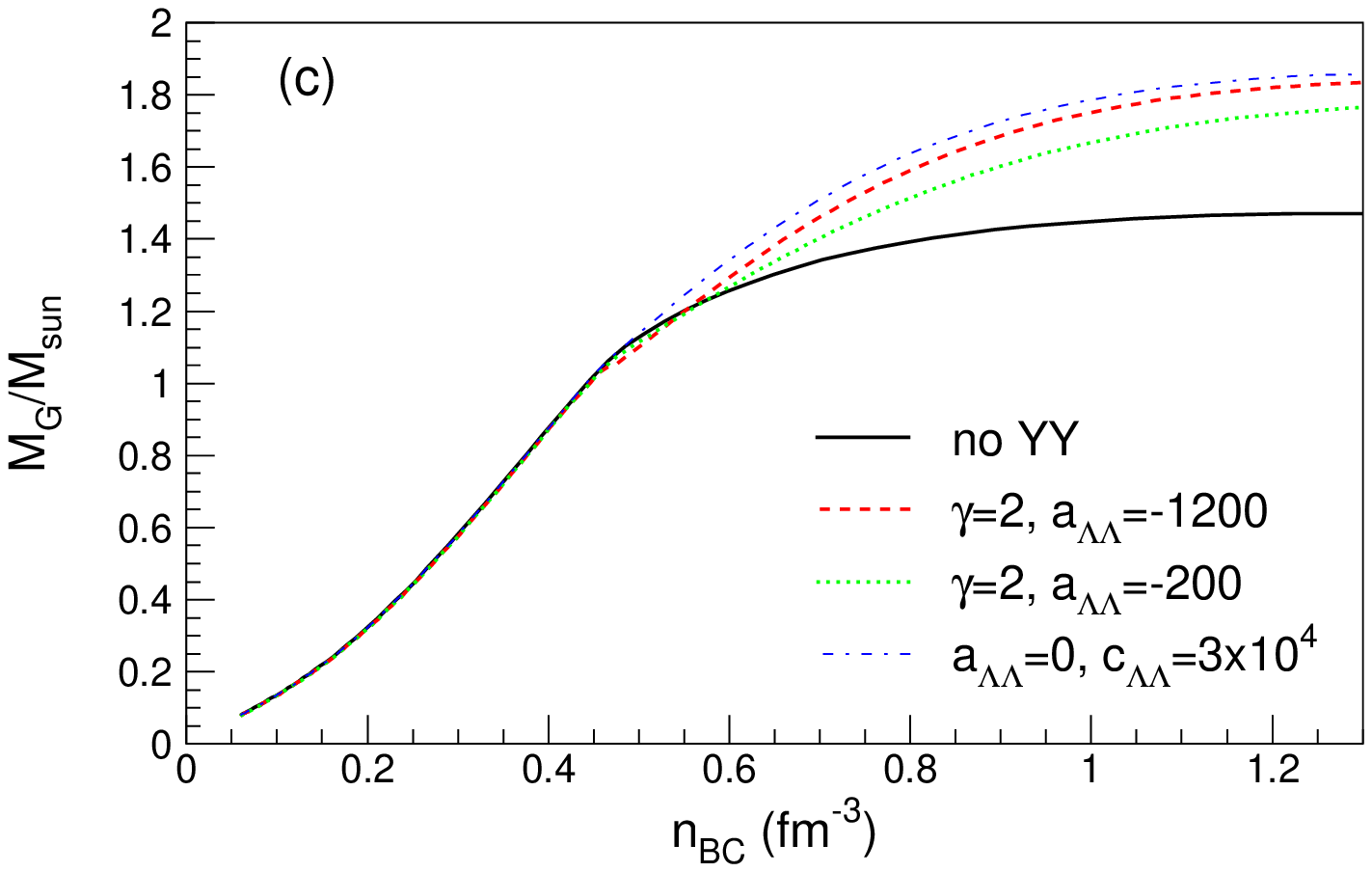}
\end{center}
\caption{(Color online)
BSL parameterization + $\Lambda$-$\Lambda$-interaction: 
(a) Limiting values of the coupling constant $a_{\Lambda \Lambda}$ for which, 
at different $\gamma$, 
the symmetric $(N,\Lambda)$-system at $T$=0
manifests strangeness driven phase transition along $\mu_S=0$;
$c_{\Lambda \Lambda}$ is fixed via the condition
$U_{\Lambda}^{(\Lambda)}(n_0/5)=-0.67 $MeV;
(b) $\Lambda$-density along the $\beta$-equilibrium path at $T$=0 
for the original BSL potential and the cases in which BSL is supplemented 
with a $\Lambda$-$\Lambda$-interaction following the functional form proposed
by Balberg and Gal~\cite{BG97} (see eq. (\ref{eq:epot_BG})) 
and obeying the condition $U_{\Lambda}^{(\Lambda)}(n_0/5)$=-0.67 MeV, 
with $\gamma$=2 and $a_{\Lambda \Lambda}$=-1200, -200 MeVfm$^3$ and,
respectively,
$\gamma$=2 and $a_{\Lambda \Lambda}$=0 and $c_{\Lambda \Lambda}$=30000 MeVfm$^{3 \gamma}$.
(c) Gravitational mass as solution of the TOV equations as a function of
central baryon number density for the cases
considered at (b).}
\label{fig:instab_Schulze_YY}
\end{figure}

We have, however, to keep in mind that this functional gives an EoS which is
much too soft and fails to reproduce $2 M_\odot$ maximum neutron star mass \cite{SR}.
This is shown in the bottom panel of Fig.~\ref{fig:instab_Schulze_YY}, which
depicts the NS mass as a function of central baryon number density.  This is
due to the lacking repulsion in the high density domain, meaning that probably
the functional is not very reliable at the densities relevant for the phase
transition. It is thus important at this point to stress that no firm
conclusion can be drawn.  It is certainly true that the phenomenological BG
form is largely arbitrary; however the description of the nucleon-hyperon
interaction in the BHF theory cannot be complete, neither.

As we have already stressed, the $Y$-$Y$ interaction cannot be neglected in
hyperonic matter.  It could well be the source of missing repulsion in
microscopic models. Due to the lack of information on this channel 
within microscopic calculations, for this
channel we will adopt the simple  polynomial form of BG, and supplement the BSL
functional,  Eq.~(\ref{eq:epot_burgio}), with it.  The upper panel of
Fig.~\ref{fig:instab_Schulze_YY} illustrates the maximum values of the
coupling constant $a_{\Lambda \Lambda}$ for which phase coexistence occurs in
symmetric $N \Lambda$ matter at various values of the stiffness parameter
$\gamma$.  The considered domains are $1.1 \leq \gamma \leq 3$ and $-1500 \leq
a_{\Lambda \Lambda} \leq -100$ MeV fm$^3$. As before, $c_{\Lambda \Lambda}$ is
obtained form the condition 
$U_{\Lambda}^{(\Lambda)}(n_0/5)=-0.67 $MeV.  One can see that, in case of
moderate $N$-$Y$-repulsion as it is the case for BSL, a phase coexistence can
still be obtained, but it requires a considerable attraction in the $Y$-$Y$
channel.

The effect of the $\Lambda$-$\Lambda$-interaction on the NS mass-central
density relation and, respectively, the $\Lambda$-hyperon abundances in
$\beta$ equilibrium is represented in the bottom and middle panels of
Fig.~\ref{fig:instab_Schulze_YY}.  The two considered
$\Lambda$-$\Lambda$-interactions correspond, respectively, to phase
coexistence ($a_{\Lambda \Lambda}$=-1200 MeV fm$^3$) and stability with
respect to phase separation ($a_{\Lambda \Lambda}$=-200 MeV fm$^3$) in
symmetric $(N,\Lambda)$-matter.  For the sake of the argument, the case of a
purely repulsive and very strong $\Lambda \Lambda$ interaction characterized
by $(\gamma=2, a_{\Lambda \Lambda}=0, c_{\Lambda \Lambda}=3 \cdot 10^4)$ is
considered, too.  We can see that employing a strongly attractive coupling at
low densities does only slightly shift the density threshold for
$\Lambda$-production with respect to a weakly attractive coupling, but
strongly enhances the equilibrium abundances just above threshold.  As we have
already noted several times, a strong attraction at low density is correlated
to a strong repulsion at high densities, leading to a relative decrease of the
$\Lambda$ abundances at higher densities with smaller values of
$a_{\Lambda\Lambda}$.  As a consequence, too, decreasing $a_{\Lambda\Lambda}$
leads to an increase of the maximum NS mass.  Though, the 2 $M_\odot$ neutron
star limit is not reached.  In conclusion, we can say that the ad-hoc
inclusion of an extra term in the BSL functional effectively accounting for
the missing $Y$-$Y$ interaction does not solve the well-known neutron star
maximum mass problem of the BHF theory.

\section{Conclusions}
\label{section:concl}

In this work, we have presented a complete study of the low temperature phase
diagram of baryonic matter including hyperonic degrees of freedom within the
phenomenological non-relativistic Balberg and Gal model \cite{BG97}. 
We have shown that
the hyperon production thresholds are systematically associated with
thermodynamic instabilities, leading to distinct first order phase
transitions.  These transitions can merge into a wide coexistence zone if the
production thresholds of different hyperonic species are sufficiently close.
As a consequence, a huge part of the phase diagram corresponds to phase
coexistence between low-strangeness and high-strangeness phases. 

In contrast to the nuclear liquid-gas phase transition which is strongly
quenched, this result is only slightly affected by adding electrons and
positrons to fulfill the charge neutrality constraint. The only effect is a
rotation of the direction of phase separation which reduces the electric
charge density component of the order parameter. The reason is that this phase
transition is driven mainly by the strangeness degree of freedom, such that
the electric charge plays only a minor role. In the latter respect, we
thus confirm the finding for the $(n,p,\Lambda,e)$-system of
Ref.~\cite{npLe} even in the presence of charged hyperons.

Along the beta-equilibrium trajectory with $\mu_L = 0$ the phase coexistence
region corresponding to the pop up of $\Lambda$- and $\Sigma$-hyperons, as
predicted by the parameterization BGI, extends over $0.3 \leq n_B \leq 0.4$
fm$^{-3}$. Physically this path is explored by neutron stars with untrapped
neutrinos. Following the study in the simple $(n,p,\Lambda,e)$-model in
Ref.~\cite{npLe}, we expect that this phase coexistence region remains at
higher temperatures and extends over density and lepton fraction $Y_L =
n_L/n_B$ domains explored by warm proto-neutron star matter. A more
quantitative analysis is left for future work. 

The possible existence of such a phase transition is strongly conditioned by
the $N$-$Y$ and $Y$-$Y$ interaction. 
In the second part of the paper we have
thus investigated on the one hand the dependence of the phase diagram on the
interaction parameters within the phenomenological BG energy density functional
and on the other hand, we have compared the results with an energy density
functional based on microscopic BHF calculations by Burgio, Schulze and
Li~\cite{BSL}.  
A complete parameter study of the energy functional would be very
cumbersome and not very illuminating, because of the huge number of
insufficiently constrained couplings.  We have therefore considered the
simplified situation of nuclear matter with $\Lambda$-hyperons. The
phase diagram of this simple model is very similar to the one obtained
including the full baryonic octet at densities below the threshold of
appearance of more massive hyperons. We therefore believe that this simple
model can give correct qualitative results for the full problem.

Both, $N$-$Y$ and $Y$-$Y$-couplings are seen to play a role in determining the
existence of an instability. Within the BG model it is shown that an
instability exists over a very large parameter domain and the two solar mass 
limit of NS is compatible with important hyperonic abundances. 
At variance, BSL is stable with respect to phase separation.
Though, phase instability can be reached when the original interaction potential
is supplemented with a phenomenological $Y$-$Y$ interaction.
We have considered both pure repulsive and attractive-repulsive potentials 
who fit the experimental data, the measurement of a positive bond
energy in double-$\Lambda$ hypernuclei.
The results show that an extra $Y$-$Y$ interaction always results in an 
enhanced maximum NS mass. 
       
In conclusion, we believe that there is no hyperon puzzle in the sense that the $2 M_{\odot}$ neutron star mass
value is well compatible with important hyperonic content and, at the same
time, the available constraints on the hyperon couplings. 

Though, whether
hyperons in neutron stars experience a phase transition is a question
which requires more constraints from the experimental side.


\acknowledgments 

This work has been partially funded by the SN2NS project 
ANR-10-BLAN-0503 and it has been supported by
NewCompstar, COST Action MP1304.
Ad. R. R acknowledges partial support from the Romanian National
Authority for Scientific Research under grants 
PN-II-ID-PCE-2011-3-0092 and PN 09 37 01 05
and kind hospitality from LPC-Caen and LUTH-Meudon.

\end{document}